\documentclass[usenatbib]{mnras}
\usepackage{graphicx}

\begin{document}

\title[Short timescale variables in the \textit{Gaia} era]{Short timescale variables in the \textit{Gaia} era: detection and characterization by structure function analysis}
\author[Roelens et al.]{Maroussia Roelens$^{1}$, 
Laurent Eyer$^{1}$, 
Nami Mowlavi$^{1}$, 
Isabelle Lecoeur-Ta\"{i}bi$^{2}$, 
\newauthor
Lorenzo Rimoldini$^{2}$,
Sergi Blanco-Cuaresma$^{1}$,  
Lovro Palaversa$^{3}$, 
Maria S\"{u}veges$^{4}$, 
\newauthor
Jonathan Charnas$^{2}$,
Thomas Wevers$^{5}$
\\
$^{1}$Observatoire de Gen\`{e}ve, D\'{e}partement d'Astronomie, Universit\'{e} de Gen\`{e}ve, Chemin des Maillettes 51, CH-1290 Versoix, Switzerland\\
$^{2}$Observatoire de Gen\`{e}ve, D\'{e}partement d'Astronomie, Universit\'{e} de Gen\`{e}ve, Chemin d'Ecogia 16, CH-1290 Versoix, Switzerland\\
$^{3}$Institute of Astronomy, University of Cambridge, Madingley Road, Cambridge CB3 OHA, United Kingdom\\
$^{4}$Max Planck Institute for Astronomy, K\"{o}nigstuhl 17, D-69117 Heidelberg, Germany\\
${^5}$Department of Astrophysics / IMAPP, Radboud University, P.O. Box 9010, NL-6500GL, Nijmegen, The Netherlands}
\date{Accepted 2017 August 12. Received 2017 July 14}

\maketitle

\begin{abstract}
We investigate the capabilities of the ESA \textit{Gaia} mission for detecting and characterizing short timescale variability, from tens of seconds to a dozen hours. We assess the efficiency of the \textit{variogram analysis}, for both detecting short timescale variability and estimating the underlying characteristic timescales from \textit{Gaia} photometry, through extensive light-curve simulations for various periodic and transient short timescale variable types.
We show that, with this approach, we can detect fast periodic variability, with amplitudes down to a few millimagnitudes, as well as some M dwarf flares and supernovae explosions, with limited contamination from longer timescale variables or constant sources. Timescale estimates from the variogram give valuable information on the rapidity of the underlying variation, which could complement timescale estimates from other methods, like Fourier-based periodograms, and be reinvested in preparation of ground-based photometric follow-up of short timescale candidates evidenced by \textit{Gaia}.
The next step will be to find new short timescale variable candidates from real \textit{Gaia} data, and to further characterize them using all the \textit{Gaia} information, including color and spectrum.
\end{abstract}

\begin{keywords}
Stars: variables: general -- surveys -- methods: data analysis -- methods: numerical -- technics: photometric
\end{keywords}

\section{Introduction}
\label{intro}

In the astronomical literature, one can find several global descriptions of variable stars. In \cite{Richter1985}, a variable star is defined as ``a star showing brightness change in the optical, over timescales of decades at most''. But what is hidden behind this quite simple concept?
Since the first reported discoveries of variable stars, such as the supernovae of 1006, 1572 and 1604 (pointed at this time as ``new very bright stars'') or the first periodic variable Mira in 1639, our understanding of stellar variability has progressed remarkably. Nowadays, hundreds of thousands of variable stars, spread all over the Hertzsprung-Russell \citep[HR,][]{Russell1914} diagram, have been identified, and classified in different categories. Various phenomena can be at the origin of the variability, be it intrinsic or extrinsic to the star, and the observed variations cover a wide timescale range, from a few tens of seconds to thousands of days. For a review of the known variable stars across the HR diagram, see \cite{EyerMowlavi2008}.

In this work, we focus on \textit{short timescale variability}, i.e. at timescales from tens of seconds to a dozen hours. A variety of astronomical objects are known to exhibit such rapid variations in their optical light-curves, including both periodic and transient variability, with amplitudes ranking from a few millimagnitudes (mmag) to a few magnitudes. 
Improving our knowledge on these specific sources would bring invaluable clues to several fields of astrophysics, e.g. stellar evolution, pulsation theories, distance estimates and physics of degenerate matter. However, only a relatively small number of short timescale variables have been identified until now. This is a direct consequence of the inherent observational constraints when dealing with such objects, first in terms of time sampling, and then in terms of photometric precision, particularly for low amplitude variability. Note nevertheless that, during the last decade, technological improvements in imaging, with the advent of Charged Coupled Devices (CCDs), made the domain of short timescale variability accessible for a large number of stars, through high cadence photometric monitoring. Hence, several space surveys, such as Kepler \citep{Borucki2010Kepler} or CoRoT \citep{Baglin2006COROT,Barge2008COROT}, and ground-based surveys, such as the Rapid Temporal Survey \citep{Ramsay2005RATS,Barclay2011RATS} and the OmegaWhite Survey \citep{Macfarlane2015OmegaWhite,Macfarlane2016OmegaWhite}, the Optical Gravitational Lensing Experiment \citep{Udalski1992OGLE}, the Palomar Transient Factory \citep{Law2009PTF} or the Catalina Sky Survey \citep{Drake2009CRTS,Drake2014Catalina}, allowed a deeper insight into this rather unexplored domain, increasing the number of known short timescale variables.

In this context, the \textit{Gaia} ESA cornerstone mission, launched in December 2013, offers a unique opportunity to drastically change the landscape. During its 5-year mission duration, \textit{Gaia} will survey more than one billion objects over the entire sky, providing micro-arcsecond astrometry, photometry down to $G \approx 20.7 \, \mathrm{mag}$ (where $G$ is the \textit{Gaia} broad-band white light magnitude) with standard errors below the mmag level for bright sources, and medium resolution spectroscopy down to $G \approx 17 \, \mathrm{mag}$ \citep{Brown2016GDR1}\footnote{For more information on \textit{Gaia} performances, please see the \textit{Gaia} webpage \url{http://www.cosmos.esa.int/web/gaia/science-performance}}. This will make a comprehensive variability search possible, including low amplitude variables. In September 2016, the first intermediate \textit{Gaia} data release (\textit{Gaia} Data Release 1, \textit{Gaia} DR1) offered a first insight to the potential of \textit{Gaia} for variability studies, with the publication of a catalogue containing 3194 Cepheids and RR Lyrae stars, of which 386 are new discoveries \citep{Clementini2016GDR1}. Besides, since July 2014, rapid analysis of daily \textit{Gaia} data deliveries by the \textit{Gaia} Science Alerts (GSA) group enabled the identification, classification and follow-up of transient sources observed by \textit{Gaia}\footnote{\textit{Gaia} Science Alerts are publicly reported on the GSA webpage \url{http://gaia.ac.uk/selected-gaia-science-alerts}} \citep{Hodgkin2013,Wyrzykowski2016}, which for example resulted in the rare discovery of an eclipsing AM CVn system \citep{Campbell2015}. Finally, recent work of \cite{Wevers2017} investigates the potential of \textit{Gaia} for investigating the fast transient sky.
In addition, the \textit{Gaia} scanning law involves fast observing cadences, with groups of nine consecutive CCD observations separated by about $4.85\,$s from each other, followed by gaps of $1\,\mathrm{h}\,46\,\mathrm{min}$ or $4\,\mathrm{h}\,14\,\mathrm{min}$ between two successive groups \citep{deBruijne2012}\footnote{For more information on the \textit{Gaia} spacecraft, instruments and observing strategy, please see  the \textit{Gaia} webpage \url{http://www.cosmos.esa.int/web/gaia/spacecraft-instruments}}. From now on, a group of nine CCD observations is referred to as a field-of-view (FoV) transit. This peculiar time sampling enables to probe stellar variability on timescales as short as a few tens of seconds. Hence, thanks to its capabilities, \textit{Gaia} is going to increase remarkably the number of identified short timescale variables, resulting in an unprecedented census of these sources in our Galaxy and beyond.

In this paper, we present our approach for the forthcoming short timescale variability analysis from \textit{Gaia} data, as part of the \textit{Gaia} Data Processing and Analysis Consortium \citep[DPAC,][]{Mignard2008DPAC}, within the Coordination Unit 7 \citep[CU7,][]{Eyer2015CU7} whose activities are dedicated to the Variability Processing. Our aim is to assess the \textit{Gaia} potential in terms of short timescale variability detection and characterization using \textit{Gaia} per-CCD photometry in the $G$ band. Our study is based on extensive light-curve simulations, for various short timescale variable types, including both periodic and transient fluctuations. Our approach is based on the \textit{variogram method}, also known as the \textit{structure function method}, which is extensively used in the fields of quasar and AGN studies, and in high-energy astrophysics \citep[see e.g.][]{Simonetti1985, MacLeod2012, Kozlowski2016}, but can also be applied to optical stellar variability\citep[see e.g.][]{Eyer1999, Sumi2005}. This variogram technique and its application to short timescale variability detection are detailed in Sect. \ref{variogramMethod}. In Sect. \ref{simulations}, we describe the two different light-curve data sets we generated for our analysis. Section \ref{analysisContinuous} summarizes the results of the short timescale variability analysis in an ideal situation, and in Sect. \ref{analysisGaiaLike} we present the expectations in the \textit{Gaia} context. Section \ref{conclu} recapitulates our conclusions.   

\section{The variogram method}
\label{variogramMethod}

\begin{figure}
\centering
\includegraphics[width=0.9\linewidth]{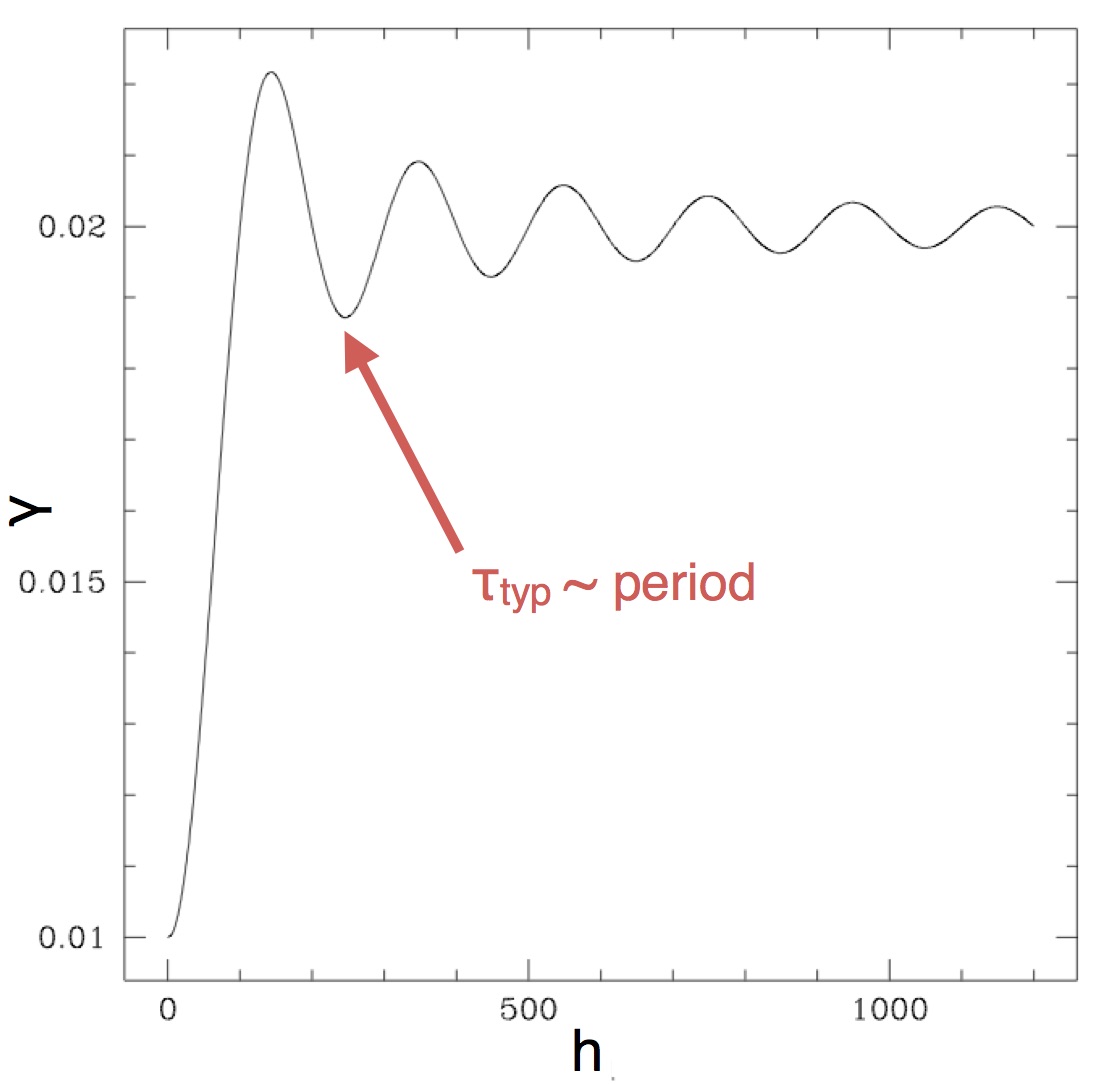}
\includegraphics[width=0.8\linewidth]{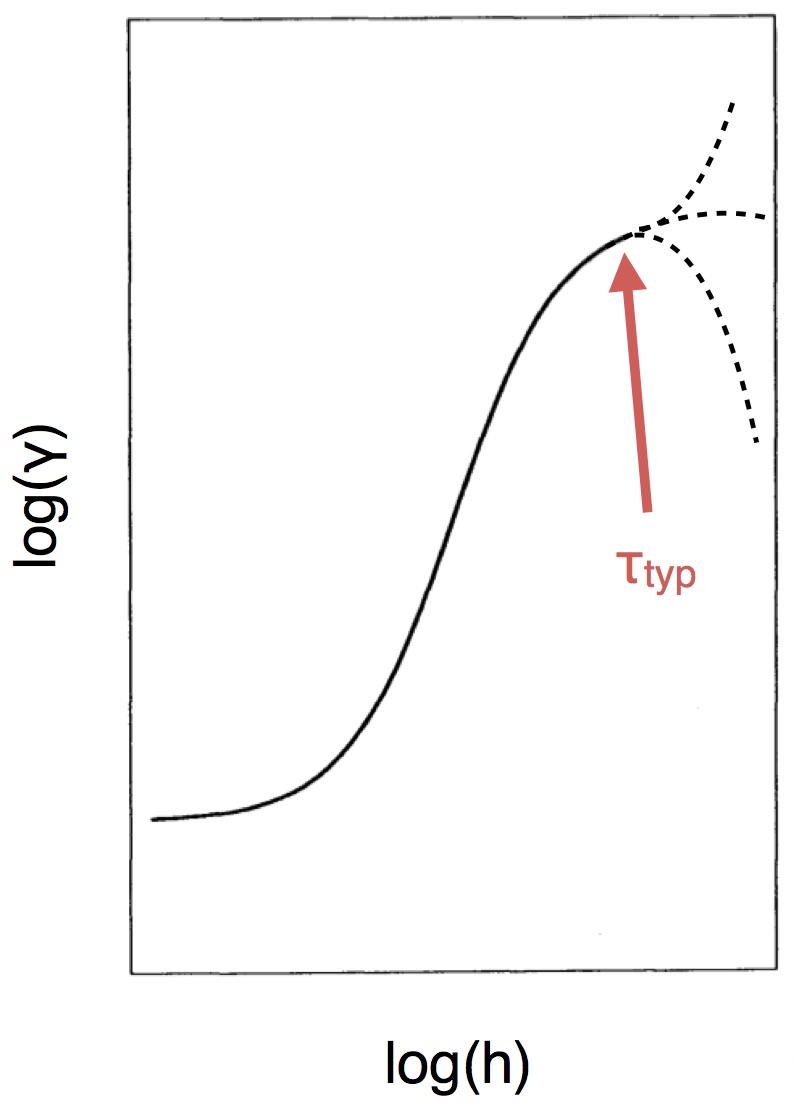}
\caption{Typical variogram plots. Top: for a periodic/pseudo-periodic variable \citep{Eyer1999}. Bottom: for a transient variable, only exploring lags up to the first structure characteristic of variability \citep[derived from][]{Hughes1992}. In each case, the feature used to estimate the typical timescale is pointed by a colored arrow.}
\label{fig:typicalVariograms}
\end{figure}

The underlying idea of our variogram approach for short timescale variability analysis is to investigate a light-curve for variability by quantifying the difference in magnitude between two measurements as function of the time lag $h$ between them. The light-curve consists in magnitudes $(m_{i})_{i=1..n}$ observed at times $(t_{i})_{i=1..n}$. The variogram value for a time lag $h$ is denoted by $\gamma(h)$. It is defined as the average of the squared difference in magnitude $(m_{j} - m_{i})^{2}$ computed on all pairs $(i, j)$ such that $| t_{j} - t_{i}|= h \pm \epsilon_{h}$ where $\epsilon_{h}$ is the tolerance accepted for grouping the pairs by time lag. This binning can be necessary, particulary in the case of uneven time sampling, to make sure to have enough pairs to compute variance for a given lag. If we note $N_{h}$ the number of such pairs,
\begin{equation}
\gamma(h) = \sum_{i > j} \frac{(m_{j} - m_{i})^{2}}{N_{h}}
\end{equation}
This formulation corresponds to the classical first order structure function as defined in \cite{Hughes1992}.
By exploring different lag values, one can build a \textit{variogram plot} (hereafter referred to as variogram) associated with the time series. This variogram provides information on how variable the considered source is, and on the variability characteristics if appropriate. Figure \ref{fig:typicalVariograms} shows the typical variograms for a periodic or pseudo-periodic variable (top), and for a transient variable (bottom). If the analyzed time series exhibits some variability, the expected features in its variogram are:
\begin{enumerate}
\item for the shortest lags, a plateau at $\gamma \sim \sigma_{noise}^{2}$ where $\sigma_{noise}$ is the measurement noise,
\item towards longer lags, an increase in the variogram values, followed by a flattening phase.
\end{enumerate}
When the underlying variation is periodic or pseudo-periodic, this flattening is followed by a succession of dips. In the case of a transient variation, the flattening can be followed by complex structures, e.g. other plateaus or a decrease in the variogram values, depending on the origin of variability (stochasticity, flares...).\\
The lags at which all those features occur in the variogram plot give indications on the variation characteristic timescales. Thus, for transient variability, the typical timescale $\tau_{typ}$ is approximately the lag at which the variogram starts flattening (see Fig. \ref{fig:typicalVariograms}). If there are more several plateaus in the variogram plot, then the considered variable has several typical timescales. For a periodic variable, $\tau_{typ}$ corresponds to the lag of the first dip after the plateauing, and gives a rough estimate of the period of the variability. Note that these interpretations suppose that the photometric time sampling involves time intervals significantly shorter than the variation timescale.

The advantage of the variogram method lies in the fact that it can handle periodic and pseudo-periodic variability as well as transient variability, and it can be used for both detecting and characterizing variable candidates. In our case, by quantifying the variability as a function of the timescale, the variogram analysis enables to isolate short timescale variable candidates from longer timescale variables (as we will see in Sect. \ref{analysisGaiaLike}), contrary to a classical $\chi^{2}$ test for example, which would combine variability at all timescales.

Suppose that a variogram similar to one of those in Fig.~\ref{fig:typicalVariograms} is derived from a \textit{Gaia} light-curve. We note $h_{k}$ the lags explored, and $\gamma(h_{k})$ the associated variogram values. The first question to answer is this one: is the considered source a true variable or not? One possible way to distinguish constant sources from variable ones is to fix a \textit{detection threshold} $\gamma_{det}$ such that:
\begin{enumerate}
\item if, for at least one value of $h_{k}$, $\gamma(h_{k}) \geq \gamma_{det}$, then the source is flagged as \textit{variable},
\item if, for all lags explored, $\gamma(h_{k}) < \gamma_{det}$, then the source is flagged as \textit{constant},
\end{enumerate}
where $\gamma_{det}$ corresponds to the variance level above which variability in the signal is considered as significant enough not to be due only to noise. Depending on the chosen $\gamma_{det}$ value, the definition of what is ``real'' variability is more or less restrictive. As we will see in Sect. \ref{analysisGaiaLike}, what drives this choice is to find an acceptable balance between optimizing the completness of the retrieved variable candidates sample and limiting the contamination from incorrect detections.
For the sources detected as variable with this criterion, we define the \textit{detection timescale} $\tau_{det}$ as the smallest lag for which $\gamma(\tau_{det}) \geq \gamma_{det}$. This detection timescale is characteristic of the underlying variability, and quantifies the variation rate in the investigated light-curve. Hence, the shorter $\tau_{det}$, the higher the averaged variation in magnitude per unit time, or in other words the higher the derivative on average. This means that, for example, a variable with relatively small amplitude but very short period or duration can be detected at the same detection timescale than a variable with longer period/duration but high amplitude.
As we are interested in short timescale variability, we complete our variability detection criterion with an additional condition on $\tau_{det}$: a source detected as variable is flagged as a \textit{short timescale candidate} only if $\tau_{det} \leq 0.5\,\mathrm{d}$. Harder limits on the detection timescale could be used to focus on the fastest phenomena observed.

Once short timescale variable candidates are identified, their variograms additionally provide estimates of their typical variation timescales as explained above. We emphasize on the fact that we do not consider the variogram as a substitute to period search methods, such as Fourier-based periodograms \citep{Deeming1975,Scargle1982}, string method \citep{Lafler1965} or analysis of variance \citep{Czerny1989}, which determine the signal period much more precisely in the case of strictly periodic variations. However, the variogram method can be complementary to the aforementioned techniques. For example, it can help distinguish the true period from aliases \citep{Eyer1999}. Besides, the variogram performs quite well for quasi periodic signals, where these period search methods usually fail.

\section{The simulated light-curves}
\label{simulations}

To evaluate the efficiency of the variogram method for detecting short timescale variables from \textit{Gaia} data, we simulate different light-curve data sets for various types of such astronomical objects. The main purpose of these simulations is to reproduce fast variability as is seen through the eyes of \textit{Gaia}. Table \ref{tab:varTypeListPeriodic} lists the periodic variable types we simulated, together with the corresponding period and amplitude ranges. Not all short-period variable types are included to this work, e.g. we do not simulate rapidly oscillating Ap (roAp) stars, nor subdwarf B variables such as EC14026 and Betsy stars. Additionally, we adopt a simplified approach, simulating each periodic light-cuve with one single period $P$ (see Sect.\ref{simuPeriodic}), hence not treating multiperiodicity. Some of the considered short timescale variables, such as $\delta$~Scuti or ZZ Ceti stars, are known to pulsate in various frequencies. However, in practice few of them have more than one mode with amplitude higher than a few tens of millimagnitudes. Hence the variogram is expected to reflect mostly the behaviour induced by the dominant pulsation mode, enabling to consider these multiperiodic sources as effectively monoperiodic in our analysis.
The simulated transient variable types and associated amplitudes and typical durations can be found in Table \ref{tab:varTypeListTransient}. Note that the word ``amplitude'' refers to the peak-to-peak amplitude in the case of periodic variability. For the transient events, the amplitude is defined as the difference between the quiescent magnitude (i.e. the magnitude of the source outside the event) and the brightest magnitude. In this work, all simulated periodic variables have periods shorter than $0.5\,$d. Among the simulated transient variables, supernovae (SNe) are not short timescale variables per se, since their duration is much longer than $1\,$d. Nevertheless, SNe can experience quite fast and significant brightening, with a variation rate of the order of $0.1\,$mag/d. Given the precision of the \textit{Gaia} $G$ photometry, if the brightening phase of a supernova is sampled by \textit{Gaia}, then we should be able to detect significant variation at the short timescale level.

\begin{figure}
\centering
\includegraphics[width=0.95\linewidth]{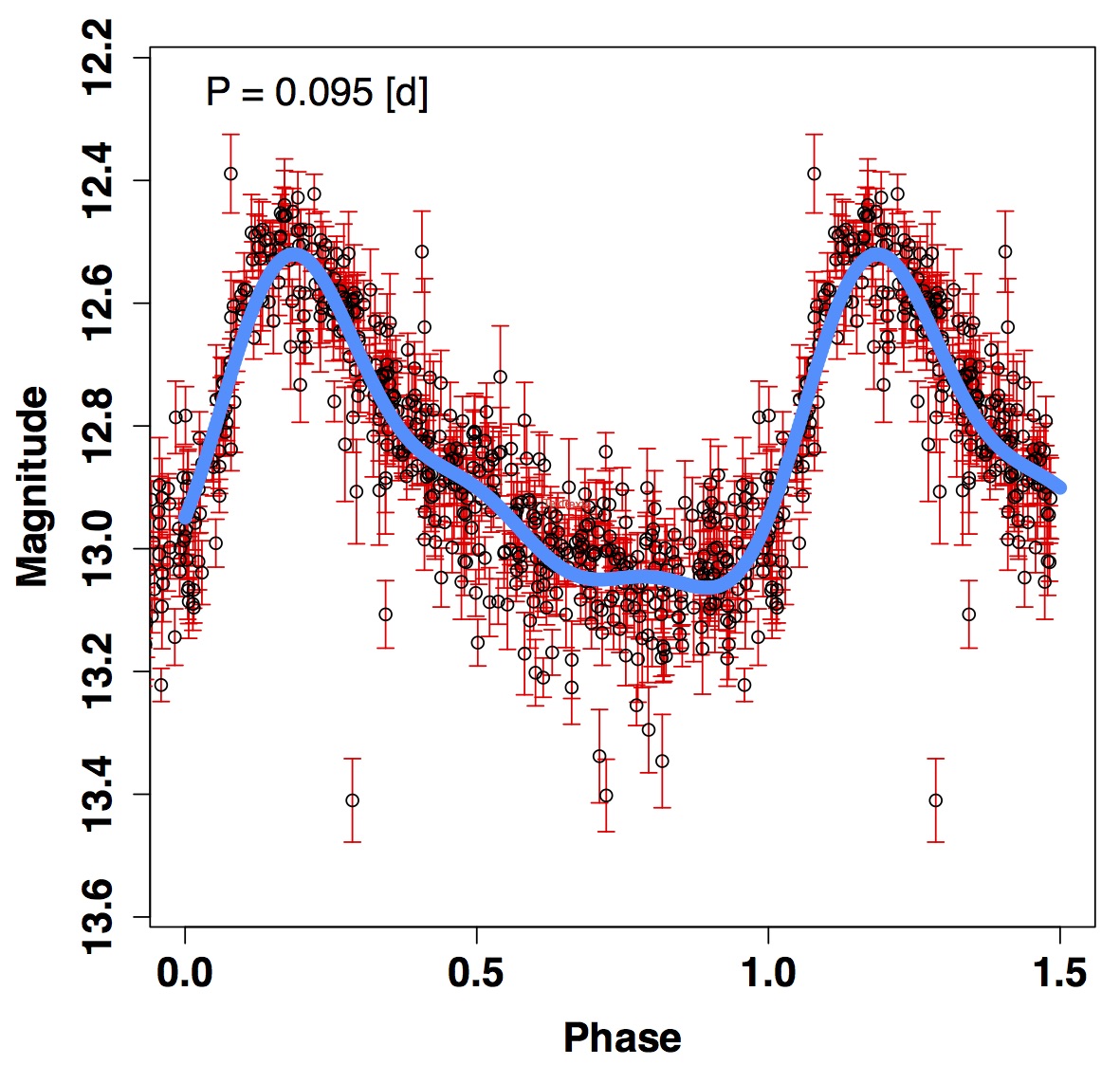}
\caption{Example of $\delta$ Scuti star template, from ASAS-3 V band measurements. The black circles with red error bars correspond to the observed ASAS light-curve. The empirical template is overplotted in blue.}
\label{fig:dsctTemplate}
\end{figure}

\begin{figure*}
\centering
\includegraphics[width=0.3\linewidth]{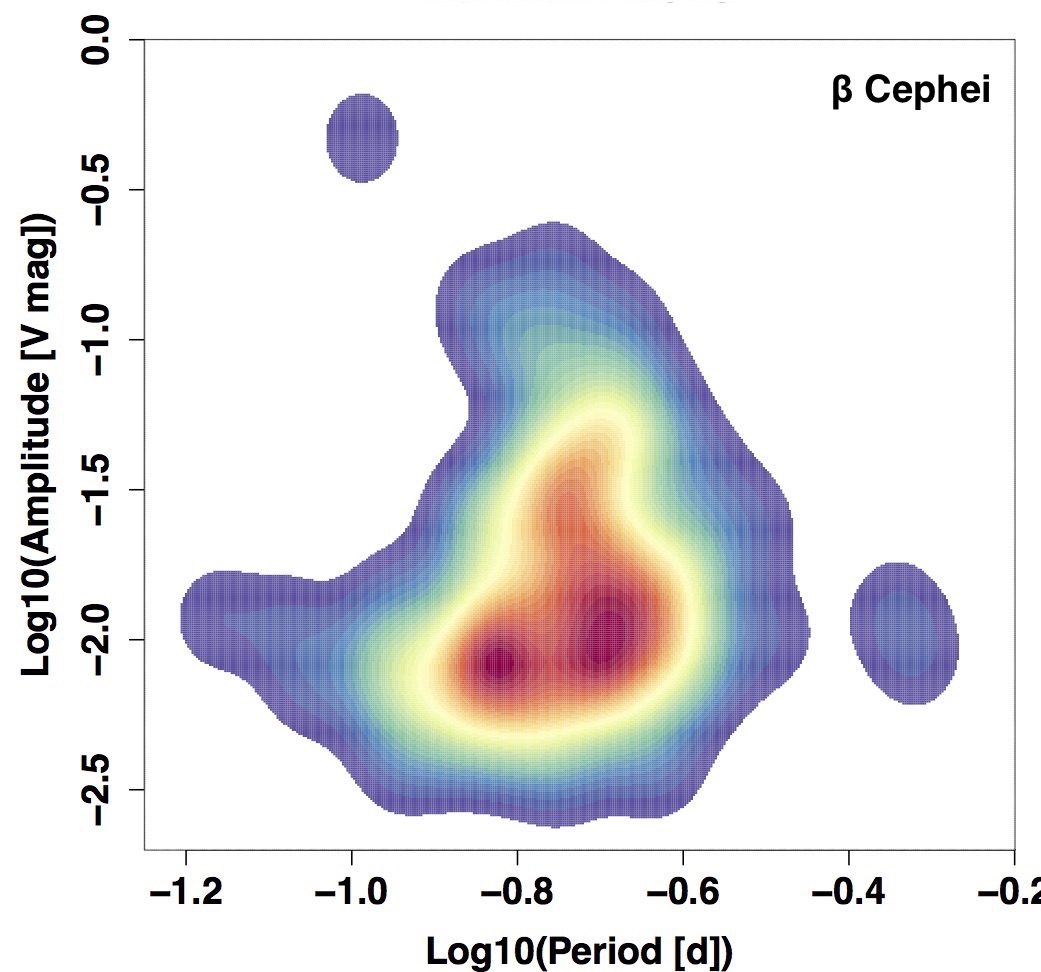}
\includegraphics[width=0.3\linewidth]{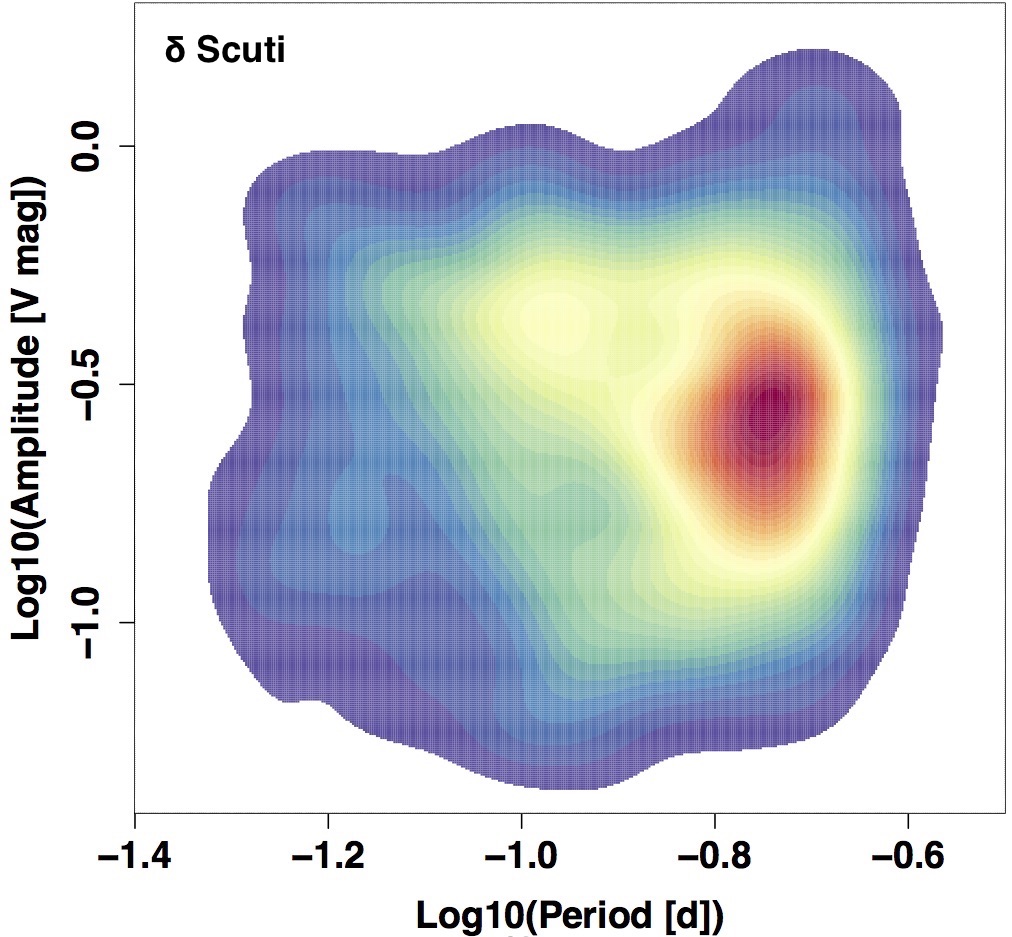}
\includegraphics[width=0.3\linewidth]{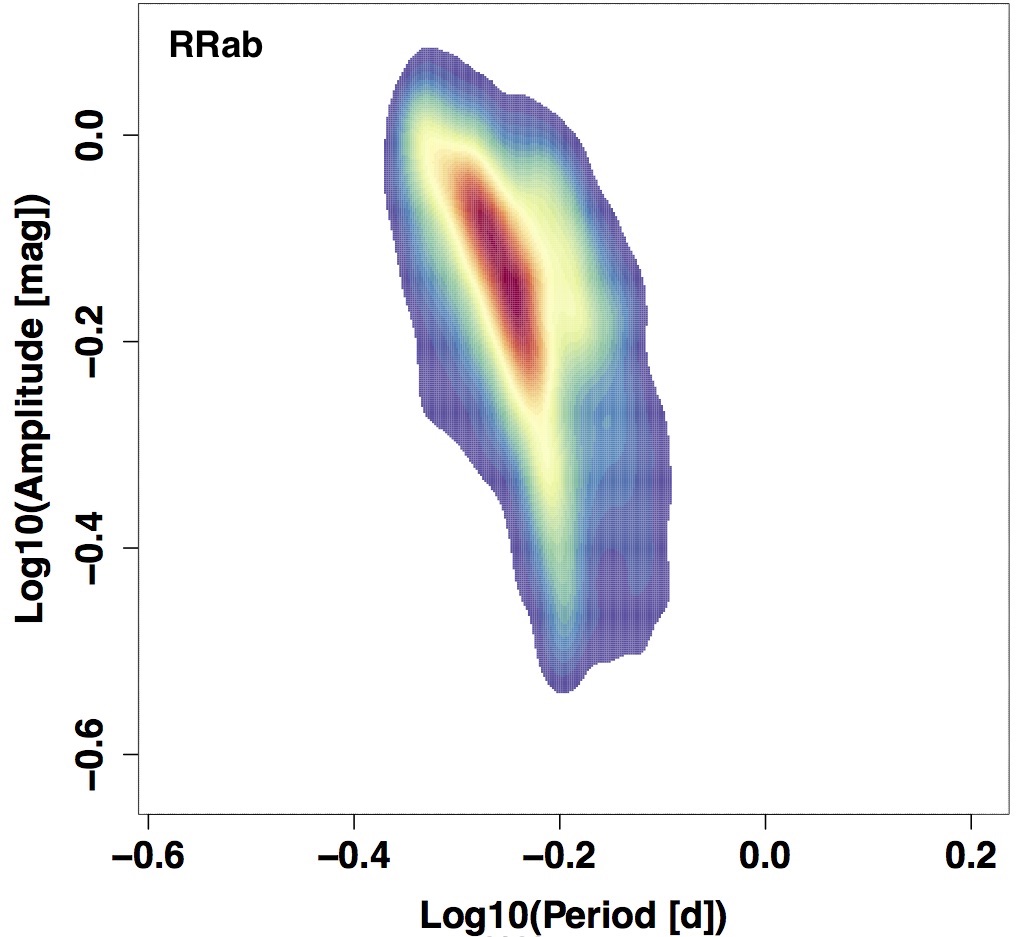}
\includegraphics[width=0.3\linewidth]{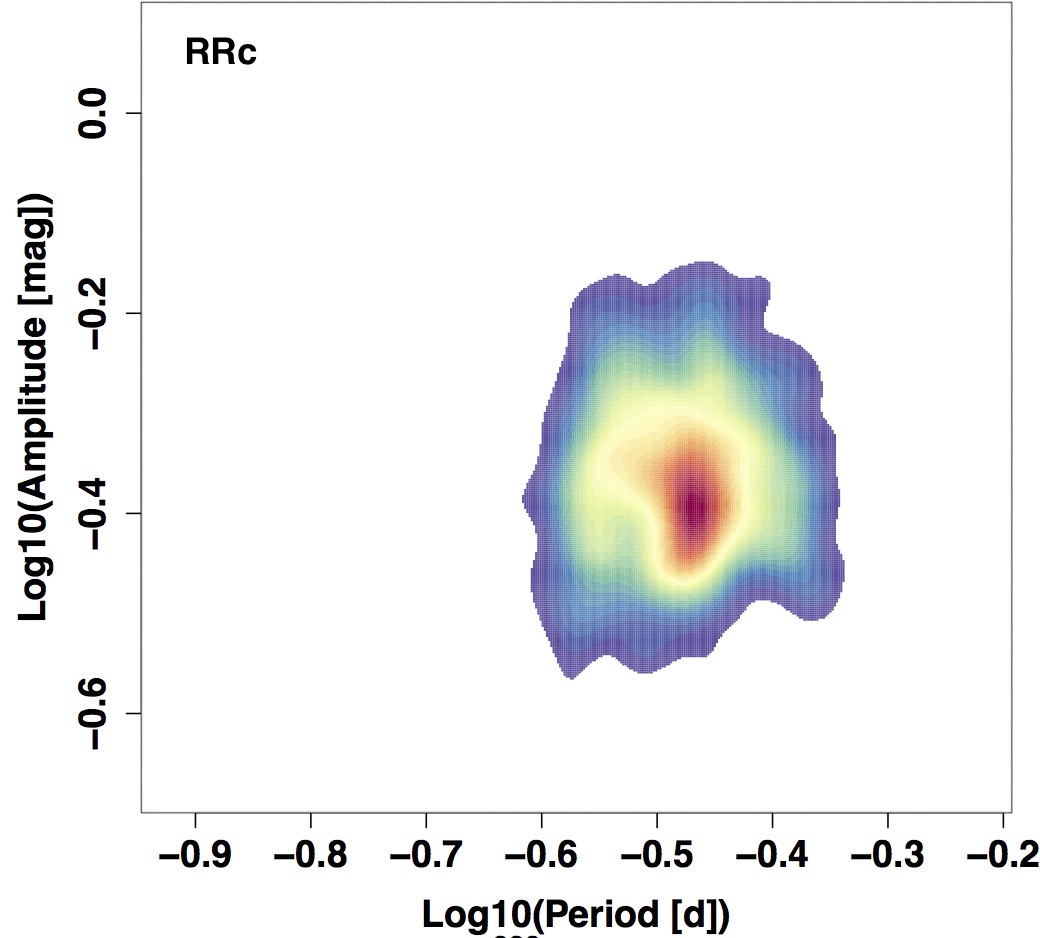}
\includegraphics[width=0.3\linewidth]{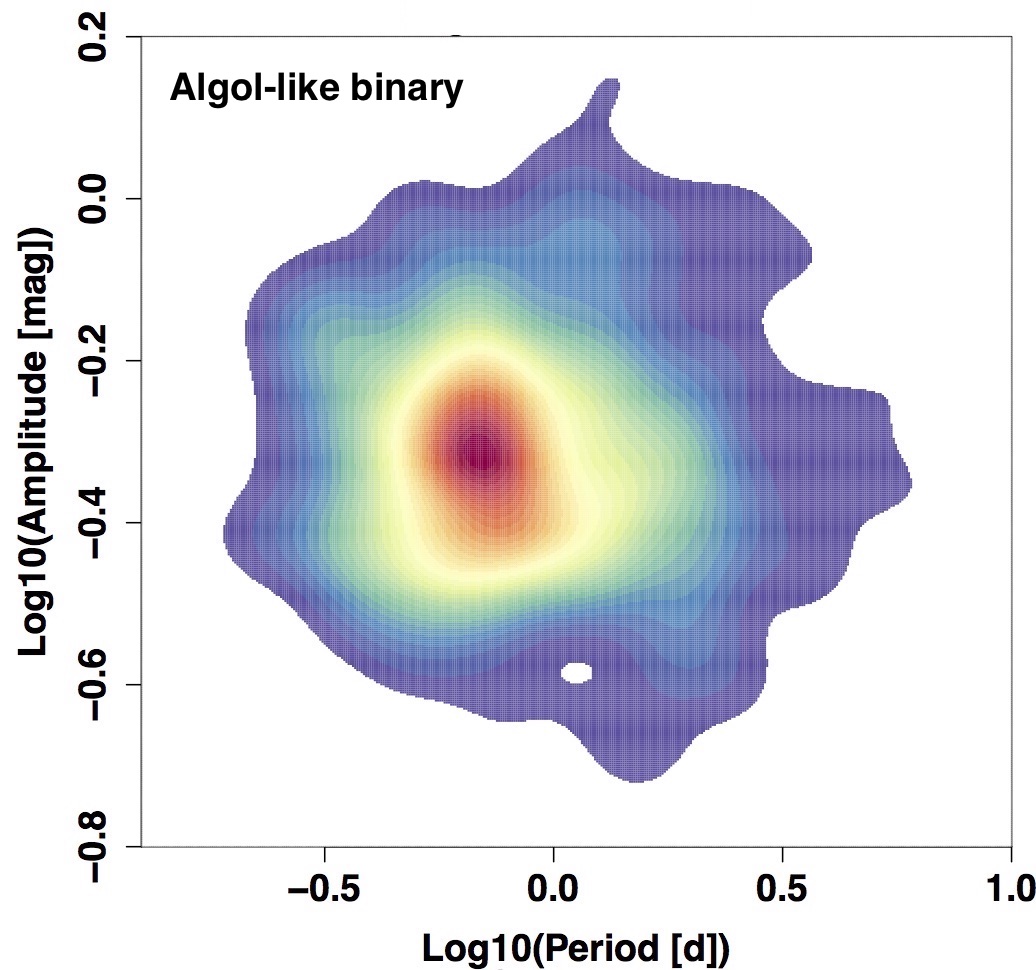}
\includegraphics[width=0.3\linewidth]{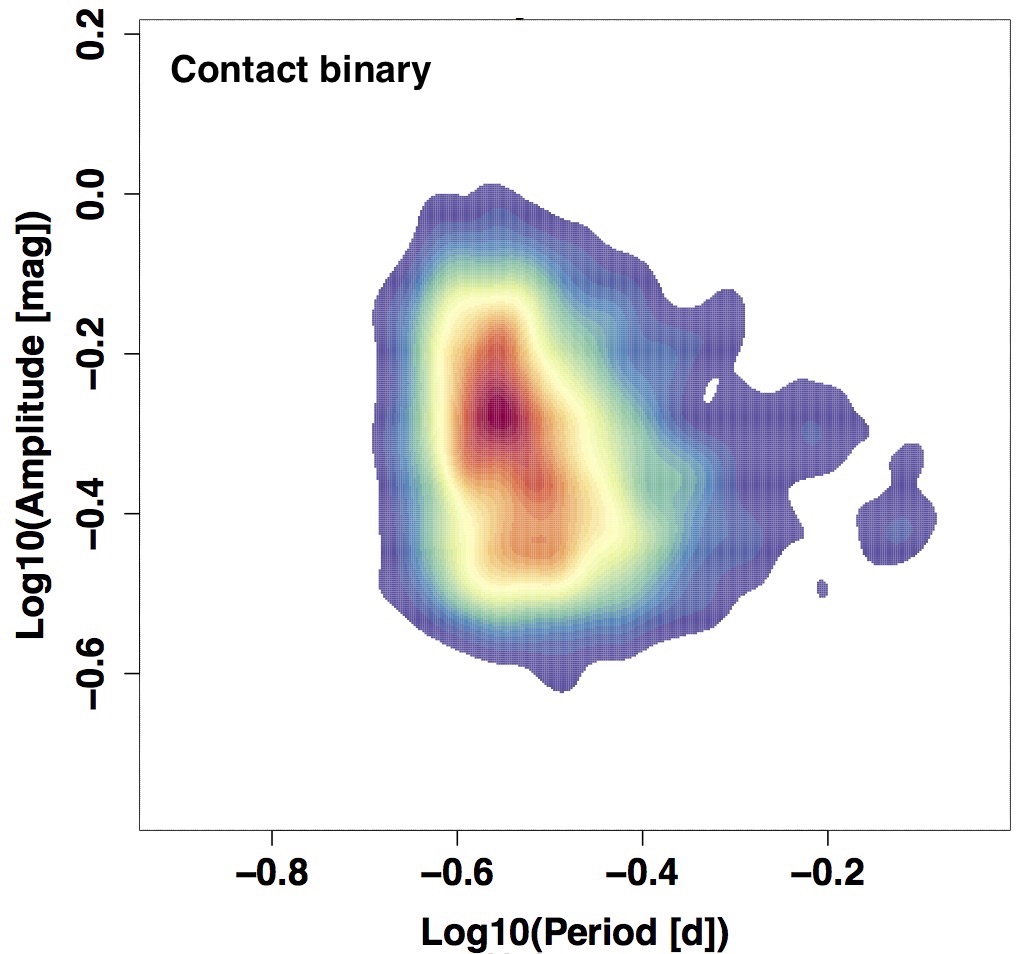}
\includegraphics[width=0.3\linewidth]{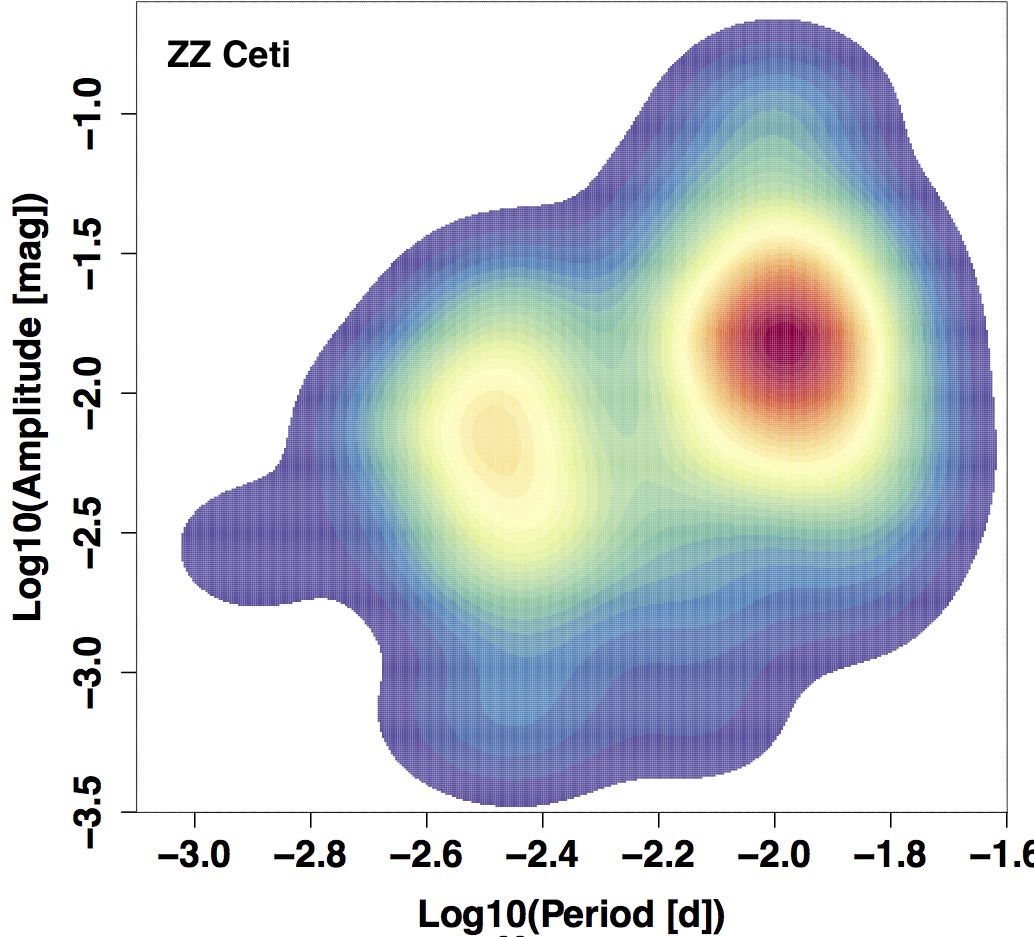}
\caption{Probability distribution in the period-amplitude diagrams. From left to right, and from top to bottom:  $\beta$ Cephei stars \citep{Pigulski2008I,Pigulski2008II}, $\delta$ Scuti stars (ASAS-3 catalogue of variable stars), RRab stars \citep{Palaversa2013}, RRc stars \citep{Palaversa2013}, Algol-like eclipsing binaries \citep{Palaversa2013}, contact eclipsing binaries \citep{Palaversa2013}, ZZ Ceti stars \citep{Mukadam2006}.}
\label{fig:allPAdistrib}
\end{figure*}

\begin{figure}
\centering
\includegraphics[width=0.9\linewidth, page=2, trim={0 0 0 1.5cm}, clip=true]{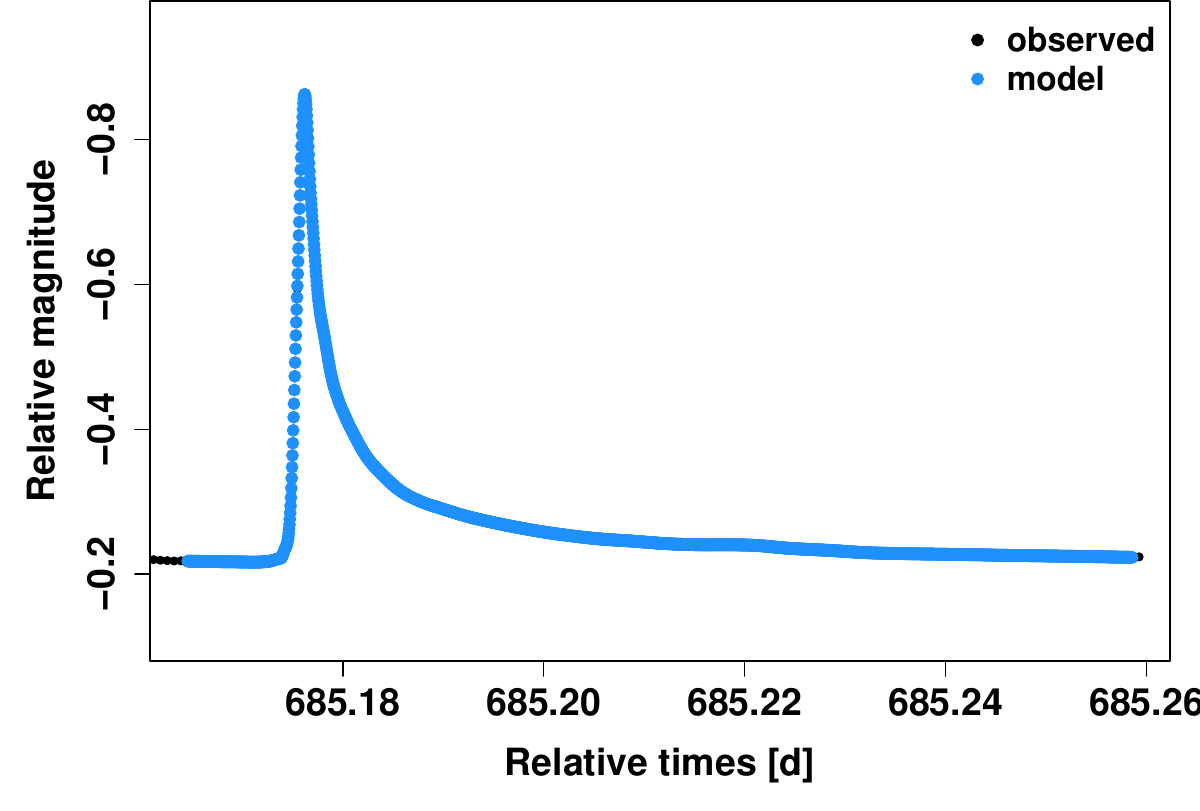}
\includegraphics[width=0.9\linewidth, page=2, trim={0 0 0 1.5cm}, clip=true]{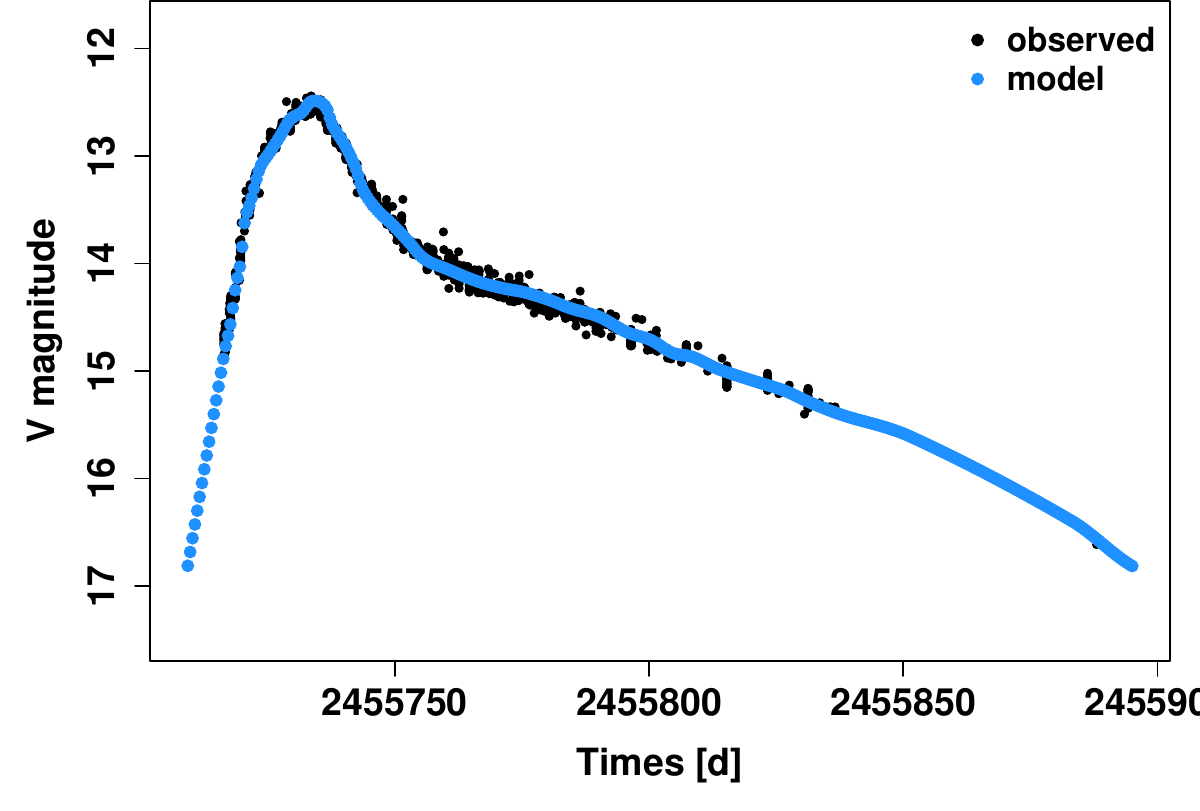}
\caption{Example of transient templates. Top: M dwarf flare template, from Kepler optical light-curve. Bottom: supernova template, from the AAVSO, with measurements in the \textit{V} band.}
\label{fig:transientTemplates}
\end{figure}

\subsection{Periodic variations}
\label{simuPeriodic}

The simulation principle we use for generating the short period light-curves is the following one. First, we build empirical, phase-folded, normalized light-curve templates, from light-curves found in the literature, for which relevant period and amplitude information are available. We retrieved 17 $\beta$ Cephei and 30 $\delta$ Scuti star templates from ASAS-3 light-curves in $V$ band \citep{Pojmanski2002}\footnote{\url{http://www.astrouw.edu.pl/asas/?page=catalogues}}. We obtained 32 RR Lyrae star and 44 eclipsing binary templates have been obtained using the LINEAR optical wide-band photometry \citep{Palaversa2013}. For ZZ Ceti star simulations, we used a set of 21 previously simulated light-curves, described in \cite{Varadi2009}. Finally, we obtained 2 AM CVn templates, one from \cite{Campbell2015} and the other from \cite{Anderson2005}, whose measurements were taken in $r$ and $g$ band respectively. Figure \ref{fig:dsctTemplate} shows an example of $\delta$~Scuti template.
The second ingredient of our recipe is the magnitude of the source to simulate. We take it uniformly between $8\,$ and $20\,$mag.
Finally, we choose the period $P$ and amplitude $A$ of the simulated variable star. To make our simulations realistic, when possible we draw the $(P, A)$ pair from empirical period-amplitude probability distributions, retrieved from existing variable star catalogs. Figure \ref{fig:allPAdistrib} represents the 2D probability distributions we used  for $\beta$ Cephei stars, $\delta$ Scuti stars, RRab stars, RRc stars, ZZ Ceti stars, Algol-like eclipsing binaries and contact eclipsing binaries. If there is not enough information in the literature, as it has been the case for AM CVn stars, we uniformly draw the period and amplitude in the appropriate ranges given in Table \ref{tab:varTypeListPeriodic}.
Once we have all these elements, we scale the phase-folded template at the simulation amplitude $A$, we generate the set of observing times, according to the appropriate time sampling over the required timespan, convert them into phases depending on the period $P$, and finally compute the corresponding magnitudes from the scaled phase-folded template.
For our analysis, we generate two different types of light-curves:
\begin{enumerate}
\item The \textit{continuous} light-curves, noiseless, with a dense and regular time sampling, over a timespan $\Delta t \sim 5P$, with about 1000 points per light-curve for AM CVn simulations, and 500 points for the other simulated variable types. The continuous data set is used to assess which periodic variable should be flagged as short timescale in an ideal situation. It comprises 100 distinct simulations for each of the 8 variable types listed in Table \ref{tab:varTypeListPeriodic}.
\item The \textit{Gaia-like} light-curves, corresponding to the same variables as in the continuous data set (same period, amplitude and magnitude), but this time with a time sampling corresponding to the expected \textit{Gaia} observation times for a random position in the sky, over a timespan $\Delta t \approx 5\,$yrs (which is the nominal duration of the \textit{Gaia} mission), and adding noise according to a magnitude-error distribution retrieved from real \textit{Gaia} data, similar to the distribution presented in fig. 6 of \cite{Eyer2017}.
\end{enumerate}
The top panels in Figs. \ref{fig:dsct84continuous} and \ref{fig:dsct84Gaialike} show two light-curves obtained for the same $\delta$ Scuti star, one simulated in the continuous way, and the other simulated in the \textit{Gaia}-like way.

\subsection{Transient variations}
\label{simuTransient}

The simulation schema applied for generating the transient light-curves is slightly different from the periodic one.
First, we build smoothed templates from light-curves found in the literature, but this time we do not apply any normalization neither in amplitude nor in duration. Contrary to what we do in the periodic case, where we use one template to simulate several time series with different periods and amplitudes, here each template is used to produce only one simulated time series for a specific event, with given amplitude and duration. Indeed, for some transients the shape of the variation depends on its peak magnitude \cite[e.g. for type Ia supernovae, ][]{Phillips1993}, hence template scaling would result in non-realistic light-curves. We retrieved 38 M dwarf flare templates from Kepler optical broad-band light-curves \citep{Davenport2014,Balona2015}, and 12 SNe templates from AAVSO\footnote{\url{http://www.aavso.org/access-data-section}} light-curves in \textit{V} band, SDSS measurements in \textit{r} band \citep{Holtzman2008,Betoule2014} and from the Harvard-Smithsonian Center for Astrophysics (CfA) archive\footnote{\url{https://www.cfa.harvard.edu/supernova/SNarchive.html}} data in \textit{V} band \citep{Bianco2014}. Figure \ref{fig:transientTemplates} shows examples of M dwarf flare (hereafter M flare) and supernova templates. For each of the 50 transient templates, we retrieve the associated increase duration $\tau_{incr}$(i.e. the time between the beginning of the event and its peak), decrease duration $\tau_{decr}$ (i.e. the time between the peak and the return to quiescent magnitude) and total duration $\tau_{tot}$ (which is the sum of the increase and decrease durations).
Then, we set the magnitude of the simulated source equal to the quiescent magnitude of the real astronomical object associated to the template used. Finally, we generate the set of observation times following the appropriate time sampling and over the required timespan, and we compute the corresponding magnitudes from our smoothed template. 
Similarly to what is done in the periodic case, from each transient template we generate two different light-curves:
\begin{enumerate}
\item The \textit{continuous} one, noiseless, with a dense and regular time sampling, over a timespan $\Delta t \sim \tau_{tot}$, and with about 1000 points per light-curve.
\item The \textit{Gaia-like} one, with a time sampling following the \textit{Gaia} scanning law for a random position in the sky, over a 5 years timespan, adding noise according to the real \textit{Gaia} magnitude-error distribution.
\end{enumerate}
The top panels in Figs. \ref{fig:Mflarecontinuous} and \ref{fig:MflareGaialike} represent the two light-curves obtained for the same M flare: the continuous one, and the \textit{Gaia}-like one. Top panels of Figs. \ref{fig:SNcontinuous} and \ref{fig:SNGaialike} show the same for a given template of supernova. Note that, for transient \textit{Gaia}-like light-curves, we ensure that at least a part of the transient event is sampled.

\begin{table*}
\centering
\caption{List of periodic short timescale variable types that are simulated}
\label{tab:varTypeListPeriodic}
\begin{tabular*}{\linewidth}{l l l l l l | l |}
\noalign{\smallskip}\hline\noalign{\smallskip}
\bf{Variable type} & \bf{Simulated} & \bf{Simulated} & \bf{Description}\\
& \bf{period range} & \bf{amplitude range [mag]} & \\
\noalign{\smallskip}\hline\noalign{\smallskip}
ZZ Ceti & 0.5 -- 25 min & $< 0.3$ & Pulsating white dwarf\\
AMCVn & 5 -- 65 min & $< 2$ & Eclipsing double (semi) degenerate system\\
$\delta$ Scuti & 28 -- 480 min & $< 0.9$ & Pulsating main sequence star\\
$\beta$ Cephei & 96 -- 480 min & $< 0.1$ & Pulsating main sequence star\\
RRab & 0.2 -- 0.5 d & 0.2 -- 2 & Pulsating horizontal branch star\\
RRc & 0.1 -- 0.5 d & 0.2 -- 2 & Pulsating horizontal branch star\\
Algol-like eclipsing binary & 0.15 -- 0.5 d & 0.2 -- 1 & Eclipsing binary of type EA\\
Contact eclipsing binary & 0.1 -- 0.5 d & 0.15 -- 0.5 & Eclipsing binary of type EB or EW\\
\noalign{\smallskip}\hline
\end{tabular*}
\end{table*}

\begin{table*}
\centering
\caption{List of transient variable types that are simulated}
\label{tab:varTypeListTransient}
\begin{tabular}{l l l l l l |}
\noalign{\smallskip}\hline\noalign{\smallskip}
\bf{Variable type} & \bf{Typical duration} & \bf{Amplitude range [mag]}\\
\noalign{\smallskip}\hline\noalign{\smallskip}
M dwarf flares & Increase $\sim 2 - 30\,$min & 0.005 -- 1.5\\
 & Decrease $\sim 30 -- 160\,$min & \\
Supernovae & Increase $\sim 15 - 165\,$d & 1.5 -- 14.5\\
 & Decrease $\sim 30 - 1500\,$d & \\
\noalign{\smallskip}\hline\noalign{\smallskip}
\end{tabular}
\end{table*}

\section{Detecting and characterizing short timescale variability in ideal cases}
\label{analysisContinuous}

\begin{figure}
\centering
\includegraphics[width=\columnwidth, page=2, trim={0 0 0 1.2cm}, clip=true]{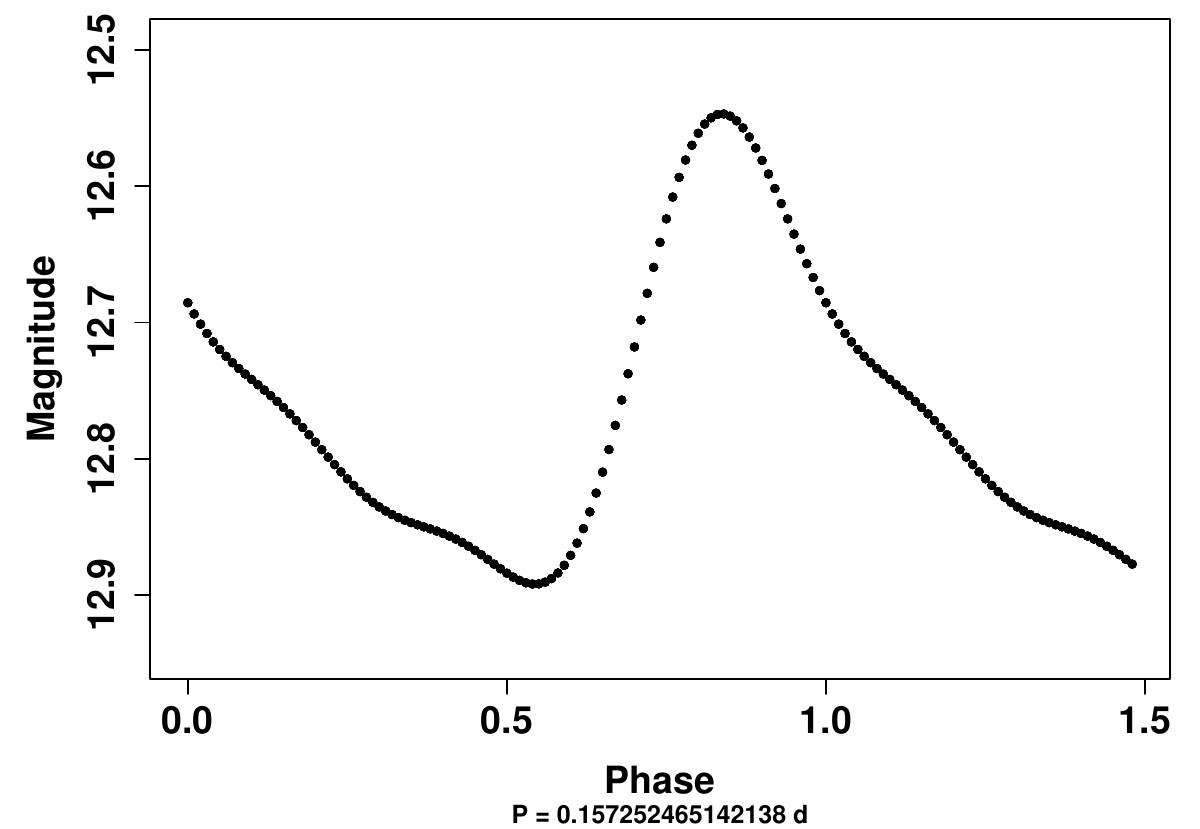}
\includegraphics[width=\columnwidth, page=3, trim={0 0 0 1.2cm}, clip=true]{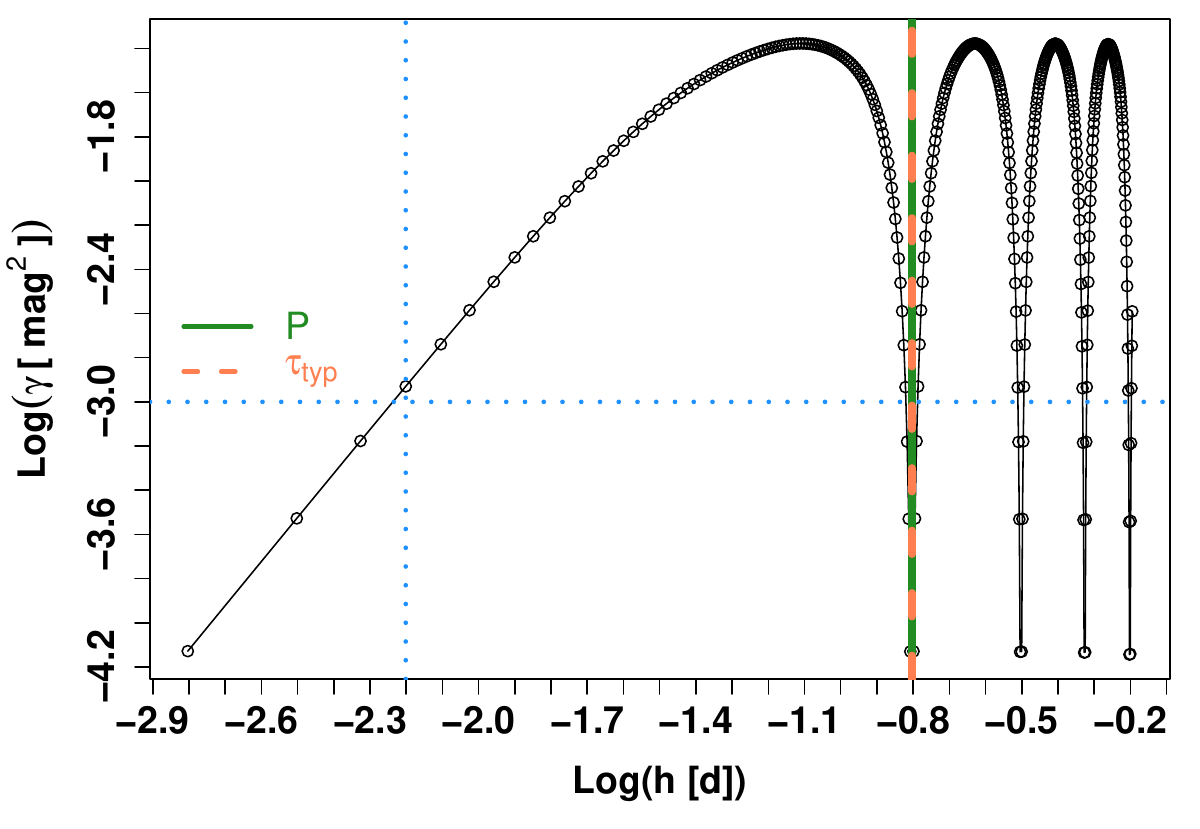}
\caption{Example of $\delta$ Scuti continuous light-curve (top), and corresponding theoretical variogram (bottom). The blue dotted lines indicate the detection threshold ($\gamma_{det} = 10^{-3} \,\mathrm{mag}^{2}$) and the associated detection timescale $\tau_{det}$. The green continuous line shows the simulation period $P$, and the orange dashed line corresponds to the typical timescale $\tau_{typ}$.}
\label{fig:dsct84continuous}
\end{figure}

\begin{figure}
\centering
\includegraphics[width=\columnwidth, page=1, trim={0 0 0 1.2cm}, clip=true]{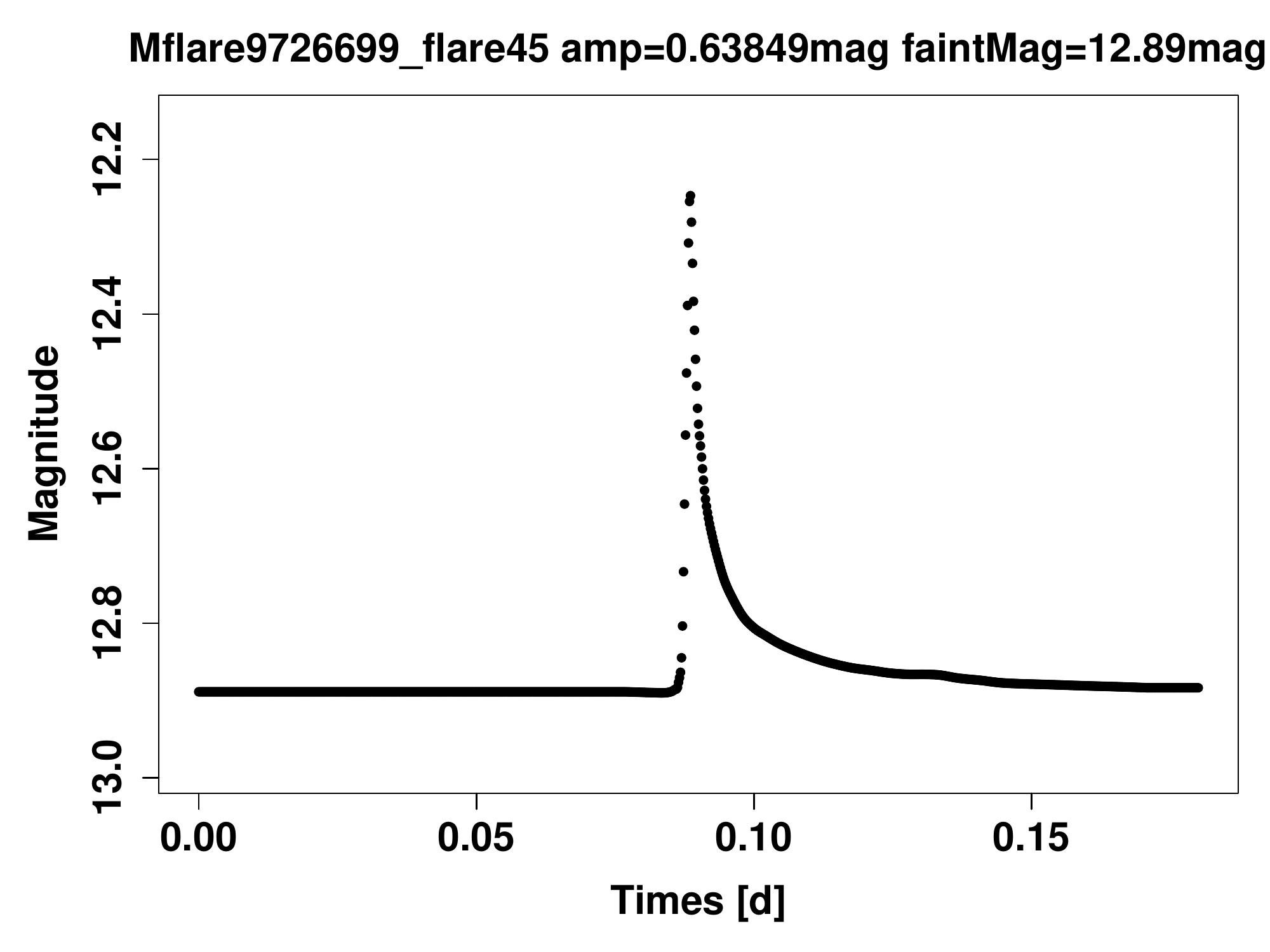}
\includegraphics[width=\columnwidth, page=2, trim={0 0 0 1.2cm}, clip=true]{Img_mnras/Mflare_9726699_flare45_variogram_R_analysis.pdf}
\caption{Example of M flare continuous light-curve (top), and corresponding theoretical variogram (bottom). The blue dashed lines indicate the detection threshold ($\gamma_{det} = 10^{-3} \,\mathrm{mag}^{2}$) and the associated detection timescale $\tau_{det}$.}
\label{fig:Mflarecontinuous}
\end{figure}

\begin{figure}
\centering
\includegraphics[width=\columnwidth, page=1, trim={0 0 0 1.2cm}, clip=true]{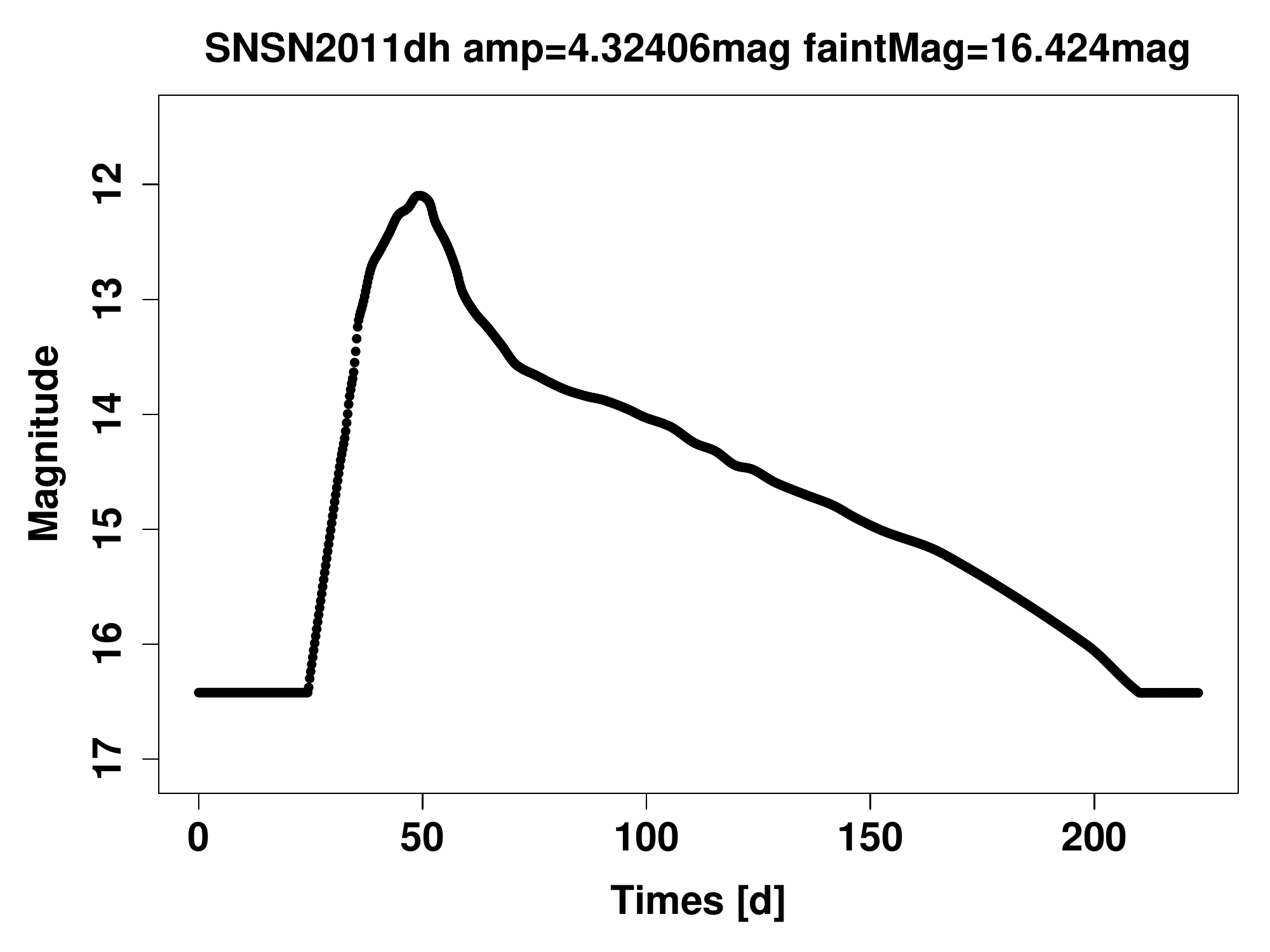}
\includegraphics[width=\columnwidth, page=2, trim={0 0 0 1.2cm}, clip=true]{Img_mnras/SN_SN2011dh_variogram_R_analysis.pdf}
\caption{Example of SN continuous light-curve (top), and corresponding theoretical variogram (bottom). The blue dashed lines indicate the detection threshold ($\gamma_{det} = 10^{-3} \,\mathrm{mag}^{2}$) and the associated detection timescale $\tau_{det}$.}
\label{fig:SNcontinuous}
\end{figure}

For each simulated continuous light-curve, we calculate the associated theoretical variogram, for the appropriate lag values defined by the underlying time sampling (i.e. explored lags are multiple of the time interval $\delta t$). Figures \ref{fig:dsct84continuous}, \ref{fig:Mflarecontinuous} and~\ref{fig:SNcontinuous} represent examples of such light-curves and variograms. Then, we apply the short timescale variability criterion described in Sect. \ref{variogramMethod}, with $\gamma_{det} = 10^{-3} \, \mathrm{mag}^{2}$ (which corresponds to a standard deviation around $0.03\,$mag), and the short timescale limit fixed at $\tau_{det} \leq 0.5\,$d. As shown in the bottom panel of Fig. \ref{fig:dsct84continuous}, the considered $\delta$ Scuti example is detected, with a detection timescale $\tau_{det} \simeq 9.1\,\mathrm{min}$. Similarly, the M flare example represented in Fig. \ref{fig:Mflarecontinuous} is detected with $\tau_{det} \simeq 1.4\,\mathrm{min}$, and the SN example of Fig. \ref{fig:SNcontinuous} is detected as well with $\tau_{det} \simeq 9.6\,\mathrm{h}$.

\subsection{Periodic variables}
\label{analysisPeriodicContinuous}

\begin{figure}
\centering
\includegraphics[width=\columnwidth, page=1, trim={0 0 0 2cm}, clip=true]{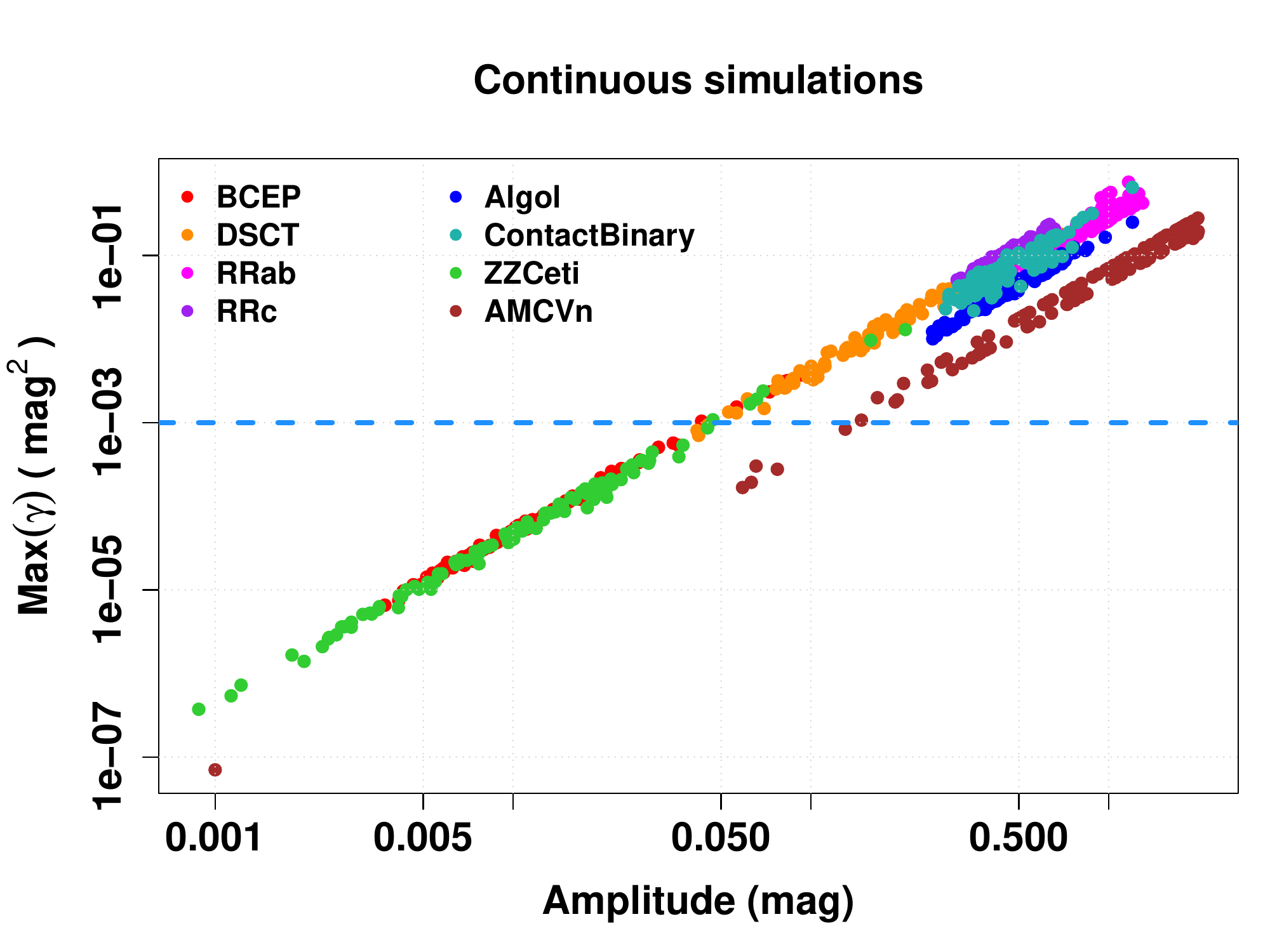}
\caption{Maximum of the variogram as function of the simulation amplitude, for the periodic variables in the continuous data set. The blue dashed line indicates the detection threshold $\gamma_{det} = 10^{-3}\,\mathrm{mag}^{2}$.}
\label{fig:maxGammaVSamplitudeContinuous}
\end{figure}

Among the 800 periodic variable simulations in the continuous data set, 603 (75.4\%) are flagged as short timescale variable with our criterion, and 197 (24.6\%) are missed. The main discriminating pattern between the flagged and missed sources is the simulation amplitude $A$. With our choice of $\gamma_{det}$, the limit amplitudes for detecting short timescale variables in an ideal noiseless case are (Fig. \ref{fig:maxGammaVSamplitudeContinuous}):
\begin{enumerate}
\item $A \ga 0.14\,$mag for AM CVn stars
\item $A \ga 0.046\,$mag for the seven other periodic variable types simulated.
\end{enumerate}
This limit is different for AM CVn stars than for the seven other simulated types, because their eclipses are very brief, so there are many more points sampling the quiescent phase than the eclipses in their time series, which results in variogram values lower than those of other variable types.

Figure \ref{fig:tauDetMosaicContinuous} represents the distributions of the detection timescale $\tau_{det}$ obtained with $\gamma_{det} = 10^{-3} \, \mathrm{mag}^{2}$ as function of the maximum variation rate in the light-curve, defined as $\max\limits_{i > j} (\vert\frac{m_{i} - m_{j}}{t_{i} - t_{j}} \vert)$, for each periodic variable type simulated. The detection timescale can be interpreted in terms of a required time sampling for source detection with the variogram method. Thus, if performing high cadence photometric monitoring of some astronomical target, and if the instrument accuracy is around $30\,$mmag, then the required imaging cadence to enable the detection of a ZZ Ceti star can be as fast as one image every $3\,$min, depending on its variation amplitude (the smaller the amplitude, the faster the cadence should be). For AM CVn stars, we see that the $\tau_{det}$ range is huge, and that this cadence for detection with variogram can go down to a few tens of seconds.

We also implement an automated method to estimate the typical timescale $\tau_{typ}$ of the variation, which in this case should approximate the period $P$ of the variable. The idea here is to identify the smallest lag whose corresponding variogram value  is greater than 80\% of the absolute maximum of the variogram, then $\tau_{typ}$ is defined by the first local minimum of $\gamma$ at lags greater than the lag of this percentaged maximum. As can be seen in Fig. \ref{fig:tauTypVSPeriodContinuous}, the detection timescale $\tau_{typ}$, deduced from the theroretical variograms of the 603 flagged sources, matches quite well their period $P$. For about 54\% of them, we have $\tau_{typ} \approx P \pm 10\,$\%. The other 46\% consist only in eclipsing binaries and AM CVn stars, and for most of them (35\% of the flagged short timescale sources) we have $\tau_{typ} \approx P/2 \pm 10\,$\%, which was quite predictable. Indeed, during one cycle of an eclipsing source, due to the presence of two minima (the primary and the secondary eclipse), pairs separated by $P/2$ have globally similar magnitudes, and therefore the variance of their magnitude difference is smaller. The remaining 11\% correspond to sources for which our timescale estimate method points a very local minimum in the variogram, due to small variations in brightness during the quiescence phase, instead of the true first dip. Note that the linear structures visible in Fig. \ref{fig:tauTypVSPeriodContinuous} are artifacts due to the simulation principle: the time step $\delta t$ is a fraction of the simulation period $P$, and the lags $h$ are multiples of $\delta t$, hence each $h \propto P$.

All in all, in an ideal situation, with the variogram method we can detect short timescale periodic variability, provided the amplitude is sufficiently large. In most cases, we can deduce from the variograms a good estimate of the underlying period, or of the semi-period for eclipsing sources. 

\subsection{Transient variables}
\label{analysisTransientContinuous}

\begin{figure*}
\centering
\includegraphics[width=0.6\linewidth, page=4]{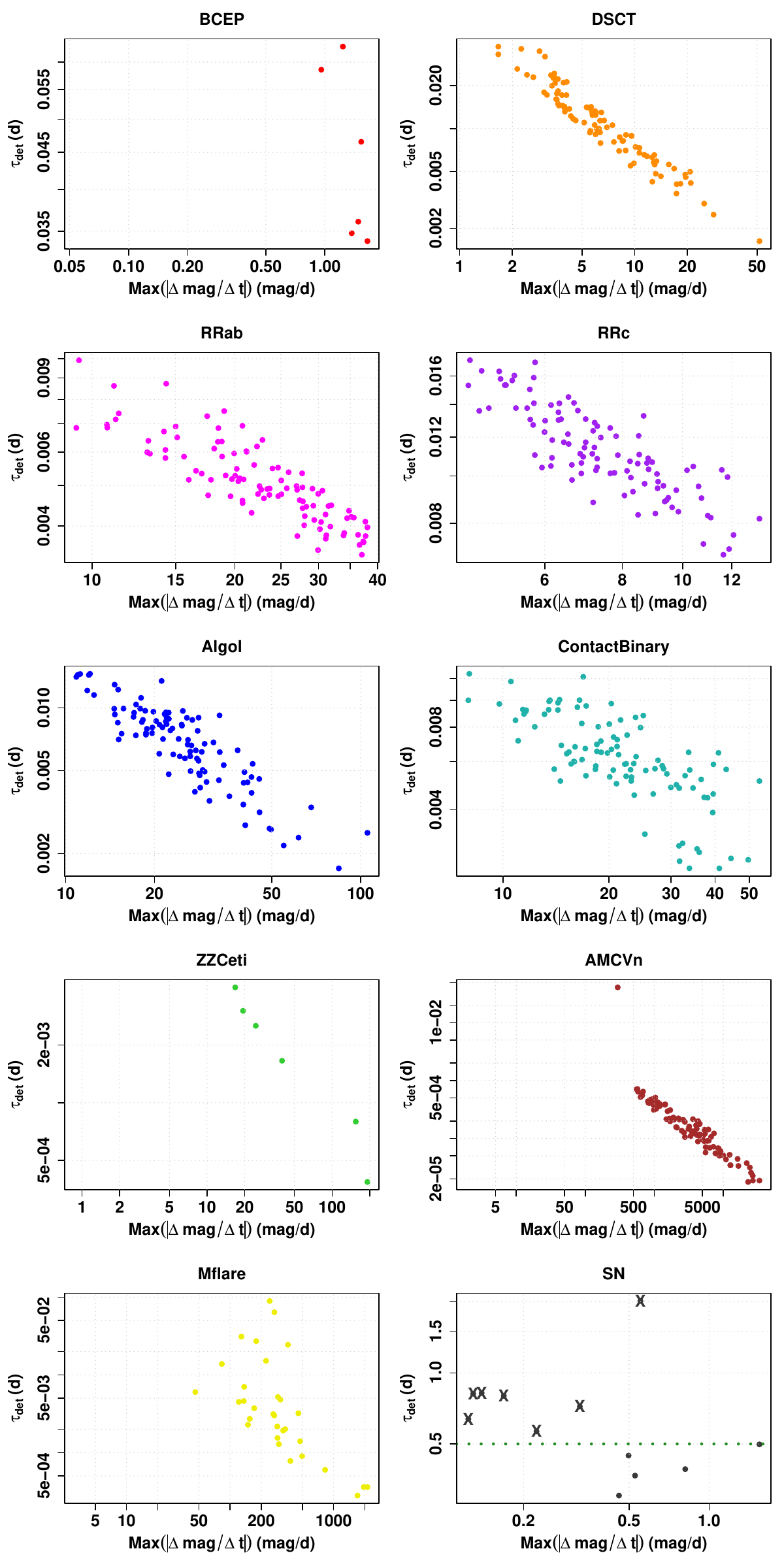}
\caption{Detection timescale as function of the maximum variation rate, for the continuous data set, for each of the 8 simulated periodic variable types. From left to right, and from top to bottom: $\beta$ Cephei, $\delta$ Scuti, RRab, RRc, Algol-like eclipsing binary, contact binary, ZZ Ceti, AM CVn, M dwarf flare, supernova. The green dashed line represents the short timescale limit that has been used, i.e. $0.5\,d$. The crosses indicate the sources that are detected ($\max(\gamma) \geq \gamma_{det}$) but not flagged as short timescale variables (i.e. $\tau_{det} > 0.5\,$d).}
\label{fig:tauDetMosaicContinuous}
\end{figure*}

\begin{figure}
\centering
\includegraphics[width=\columnwidth, page=10, trim={0 0 0 2cm}, clip=true]{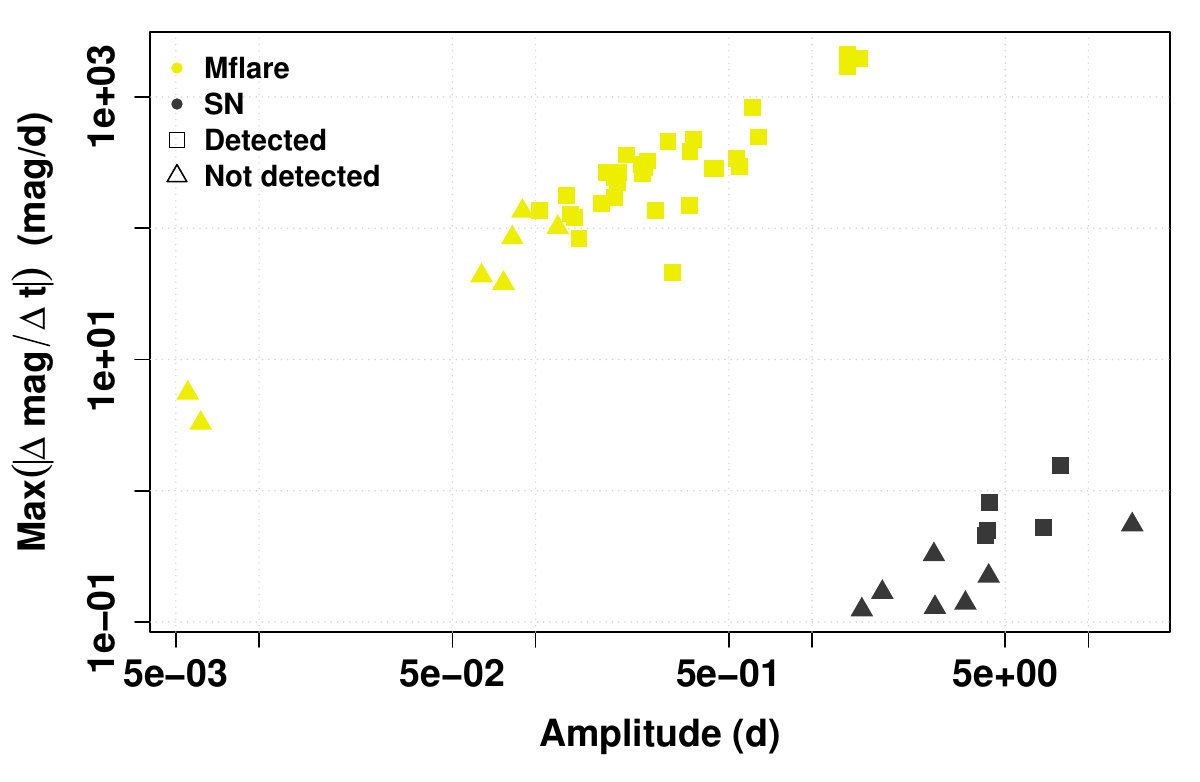}
\caption{Maximum variation rate as function of the simulation amplitude, for the transient simulations in the continuous data set.}
\label{fig:maxVariationRateVSamplitudeContinuous}
\end{figure}

\begin{figure}
\centering
\includegraphics[width=\columnwidth, page=2, trim={0 0 0 2cm}, clip=true]{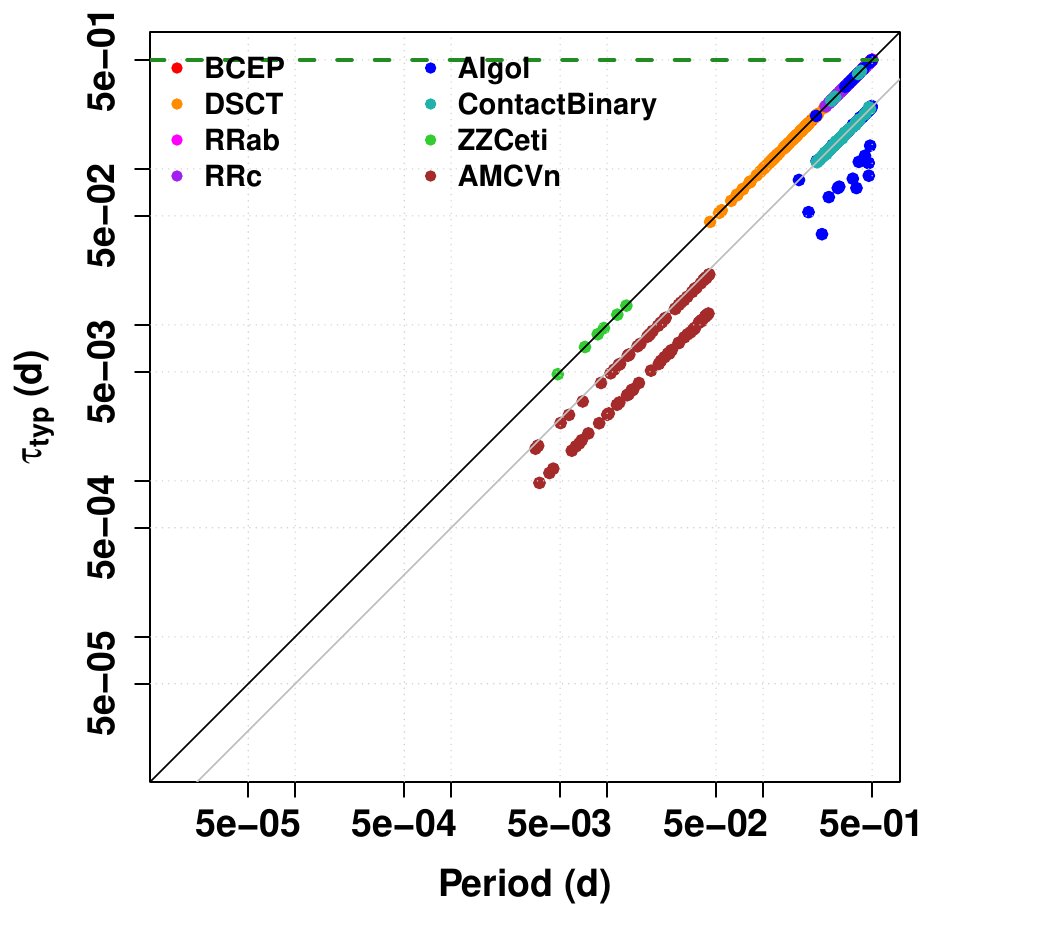}
\caption{Typical timescale $\tau_{typ}$ as function of the simulation period, for the periodic continuous simulations flagged as short timescale candidates with $\gamma_{det} = 10^{-3}\,\mathrm{mag}^{2}$. The black line indicates the first bissector, the grey line corresponds to $\tau_{typ} = P/2$, and the green dashed line corresponds to the short timescale limit of $0.5\,$d.}
\label{fig:tauTypVSPeriodContinuous}
\end{figure}

Among the 50 transient simulations in the continuous data set, 31 of the 38 simulated M dwarf flares are flagged as short timescale variables, as well as 5 of the 12 simulated supernovae. As one can see in Fig. \ref{fig:maxVariationRateVSamplitudeContinuous}, the missed M flares have the smallest amplitudes, and they are not detected at all. Typically, for this variable type and with the chosen $\gamma_{det}$, the limit amplitude for detection is $A \ga 0.12\,$mag. Furthermore, the missed supernovae are associated with smaller maximum variation rates, with a limit value of $\max\limits_{i > j} (\vert\frac{m_{i} - m_{j}}{t_{i} - t_{j}} \vert) \sim 0.15\,$mag/d. In this case, simulated sources are detected, but because their detection timescale is longer than $0.5\,$d, they are not flagged as short timescale variables. As explained in Sect. \ref{analysisPeriodicContinuous}, we can interpret the detection timescale as a prescription for future photometric follow-up. Thus, to detect an M dwarf flare from the ground with an instrument whose accuracy is around $30\,$mmag, the required observing cadence can be as high as every $4\,$min (Fig. \ref{fig:tauDetMosaicContinuous}).

\begin{figure}
\centering
\includegraphics[width=\columnwidth, page=5, trim={0 0 0 2cm}, clip=true]{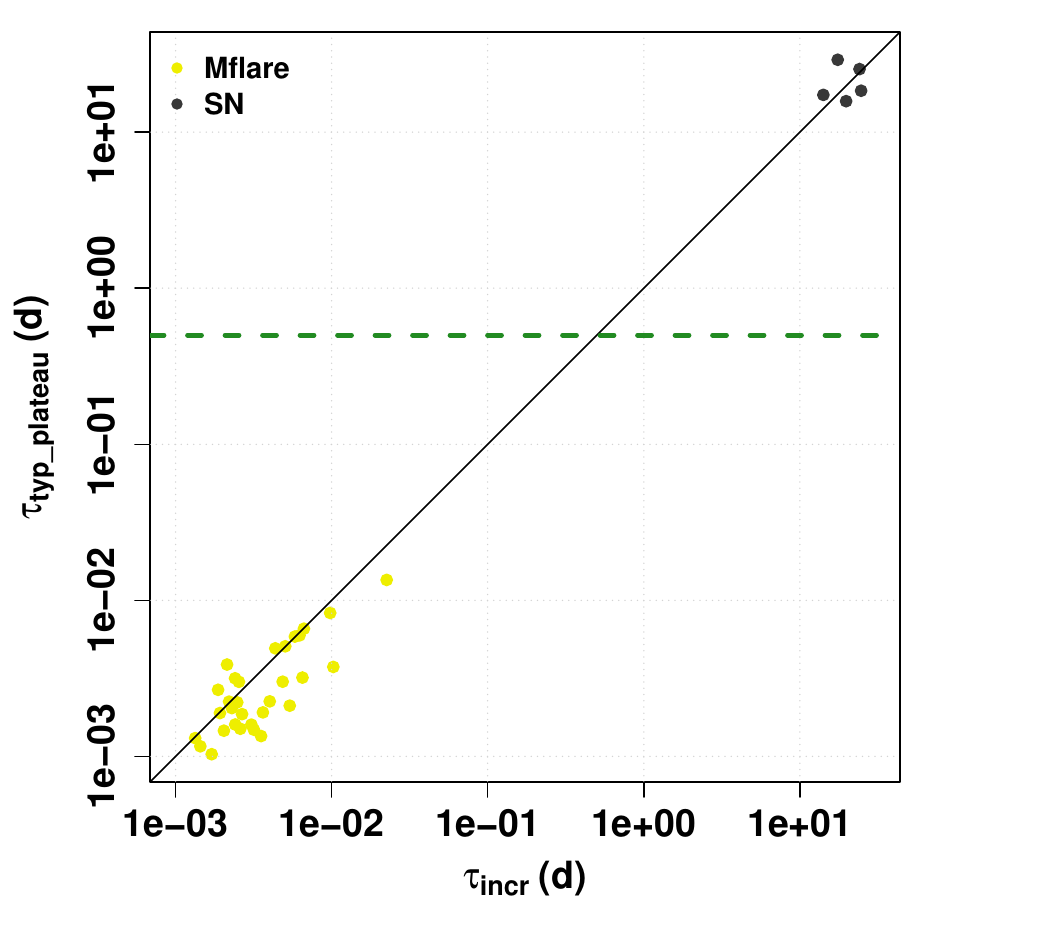}
\includegraphics[width=\columnwidth, page=6, trim={0 0 0 2cm}, clip=true]{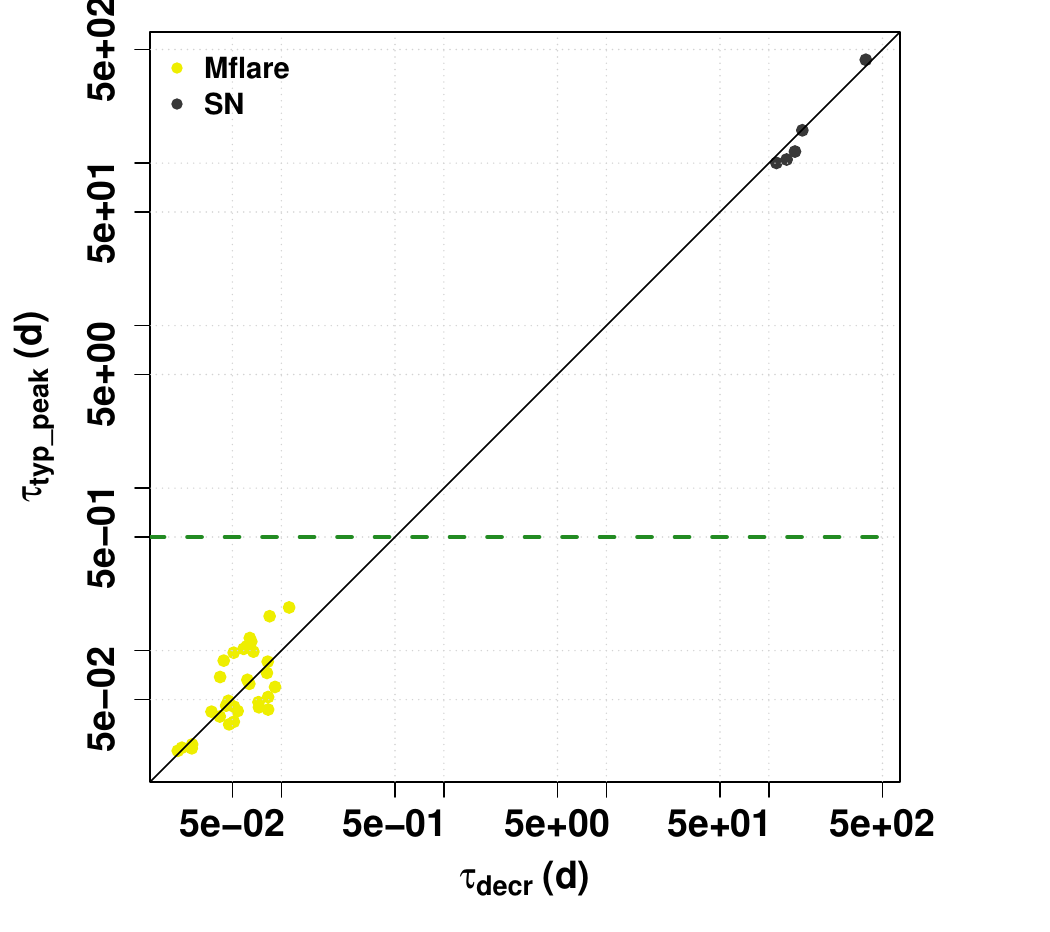}
\includegraphics[width=\columnwidth, page=7, trim={0 0 0 2cm}, clip=true]{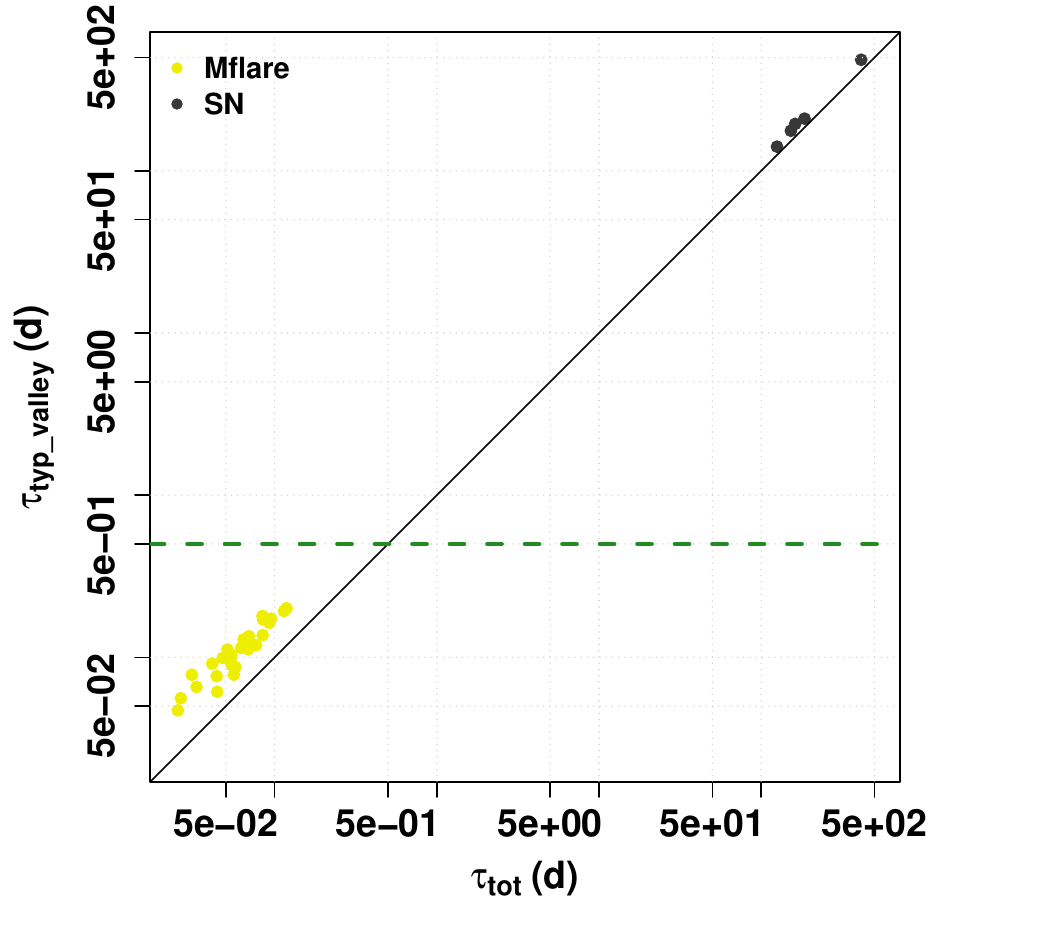}
\caption{Typical timescales as function of the event durations, for the transient continuous simulations flagged as short timescale candidates with $\gamma_{det} = 10^{-3}\,\mathrm{mag}^{2}$. The black line indicates the first bissector, and the green dashed line corresponds to the short timescale limit of $0.5\,$d.}
\label{fig:tauTypVSDurationContinuous}
\end{figure}

As for periodic variables, we estimate the different typical timescales revealed by the theoretical variograms of the flagged transients. From visual inspection of these variograms, we note the following structures:
\begin{enumerate}
\item A flatening in the variogram plot, at shorter lags, with associated timescale $\tau_{typ,plateau}$ (for the M dwarf flare example showed in Fig. \ref{fig:Mflarecontinuous}, it occurs around $10^{-2.6}\,$d$\sim 3\,$min),
\item A peak towards longer lags, occuring at timescale $\tau_{typ,peak}$
\item A valley after the aforementioned peak, starting at timescale $\tau_{typ,valley}$
\end{enumerate}
These typical timescales trace the increase, decrease and total duration of the transient, respectively. As shown in Fig.~\ref{fig:tauTypVSDurationContinuous}, the values obtained match quite well the event durations, though $\tau_{typ,valley}$ slightly overestimates the total duration.

Similarly to Sect. \ref{analysisPeriodicContinuous}, we see that in the ideal case, the variogram method enables to detect fast transient events such as M dwarf flares, provided their amplitude is higher than $\sim 0.12\,$mag, as well as some supernovae. Moreover, the typical timescale estimates retrieved from theoretical variograms recover quite well the characteristic durations of the simulated transient events.

\section{Detecting and characterizing short timescale variability in \textit{Gaia}-like observations}
\label{analysisGaiaLike}

In Sect. \ref{analysisContinuous}, we investigated the ability of the variogram method for detecting and characterizing short timescale variability, from noiseless, regularly and well sampled light-curves of various variable stars. We showed that, in an ideal situation, with the detection criterion we applied, we should detect:
\begin{enumerate}
\item $\beta$ Cephei, $\delta$ Scuti, RR lyrae, ZZ Ceti stars and eclipsing binaries with $A \ga 0.046\,$mag,
\item AM CVn stars with $A \ga 0.14\,$mag,
\item M dwarf flares with $A \ga 0.12\,$mag,
\item some supernovae, provided they have high enough variation rate.
\end{enumerate}
But what happens if we move to the \textit{Gaia}-like context? Similarly to what we did with the continuous data set, we compute observational variograms associated with each simulated short timescale variable \textit{Gaia}-like light-curve. This time, the explored lags are defined by the \textit{Gaia} scanning law, i.e. the time intervals between CCD measurements ($4.85\,$s, $9.7\,$s, $14.6\,$s, $19.4\,$s, $24.3\,$s, $29.2\,$s, $34\,$s and $38.8\,$s), and those between the different FoV transits ($1\,$h$\,46\,$min, $4\,$h$\,14\,$min, $6\,$h, $7\,$h$\,46\,$min, etc), up to $h \approx 1.5\,$d. Note that no lag can be explored from about $40\,$s to $1\,$h$\,46\,$min, which may have consequences on the detectability and characteristic timescales ($\tau_{det}$ and $\tau_{typ}$) of some sources.
For a given lag $h$, the observational variogram value is now computed on all pairs $(i,j)$ such that $\vert t_{j} - t_{i} \vert = h \pm \epsilon_{h}$. The tolerance on time lag $\epsilon_{h}$ is of $0.9\,$s for the lags smaller than $40\,$s, and of $14.4\,$min for the lags between $1\,$h$\,46\,$min and $1.5\,$d. Hence, the variogram values computed at $h = 1\,$h$46\,$min for example, group all the pairs between the 9 CCDs of one field-of-view and the 9 CCDs of the following field-of-view (e.g. CCD1 of FoV1 -- CCD1 of FoV2, CCD1 of FoV1 -- CCD2 of FoV2, etc).

Ensuring that short timescale variables can be efficiently detected by \textit{Gaia} is necessary, but not sufficient. We also have to make sure that our method limits the number of incorrect detections, for true short timescale variable candidates not to be mixed with an overflow of spuriously detected constant sources (hereafter false positives) nor with too many sources exhibiting variability at timescales longer than half a day (e.g. longer period pulsating stars such as Cepheid or Mira stars). To assess this contamination, we complete our \textit{Gaia}-like data set with:
\begin{enumerate}
\item 1000 simulations of constant star \textit{Gaia}-like light-curves, with magnitudes between $8\,$mag and $20\,$mag,
\item 100 monoperiodic sinusoidal \textit{Gaia}-like light-curves (hereafter long period variables), with periods between $10$ and $100\,$d, amplitudes between $1\,$mmag and $1\,$mag, and magnitude between $8$ and $20\,$mag,
\end{enumerate}
and perform variogram analysis for each of them, similarly to what is done for short timescale variables. 

\begin{figure}
\centering
\includegraphics[width=\columnwidth, page=2, trim={0 0 0 1.2cm}, clip=true]{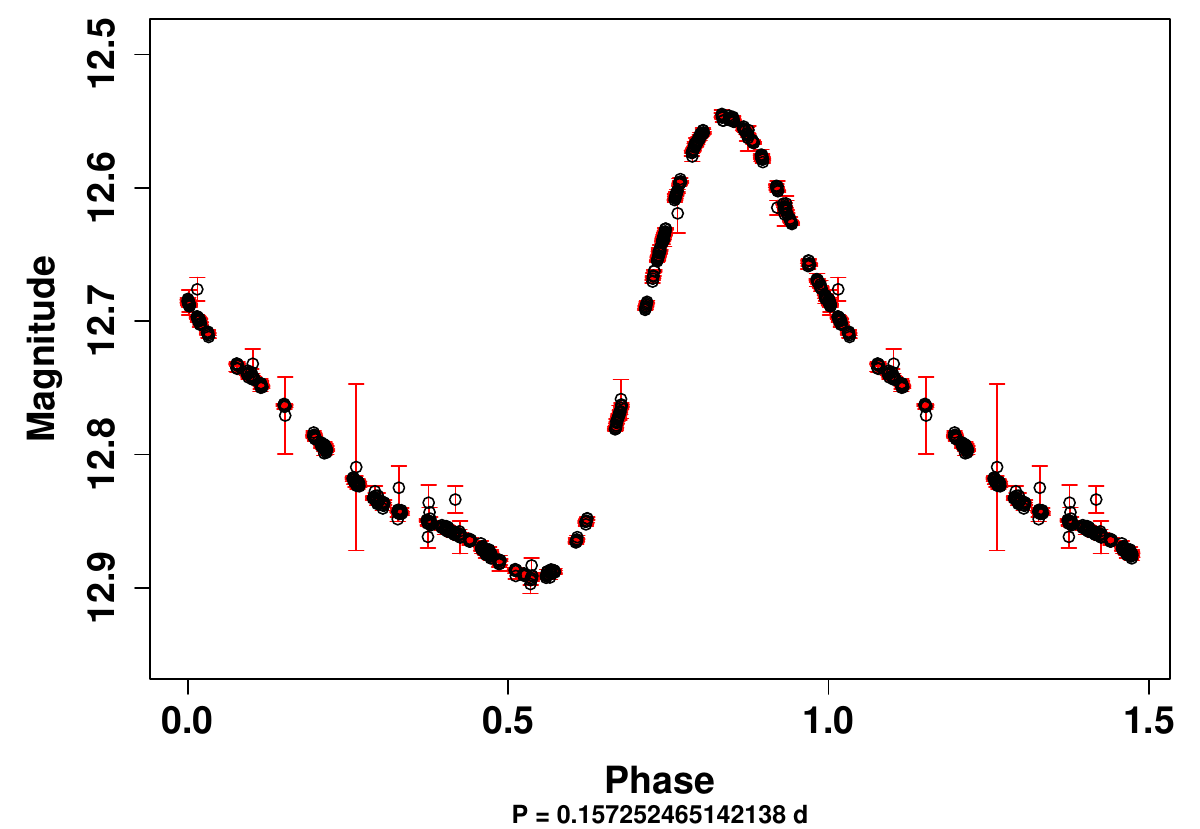}
\includegraphics[width=\columnwidth, page=3, trim={0 0 0 1.2cm}, clip=true]{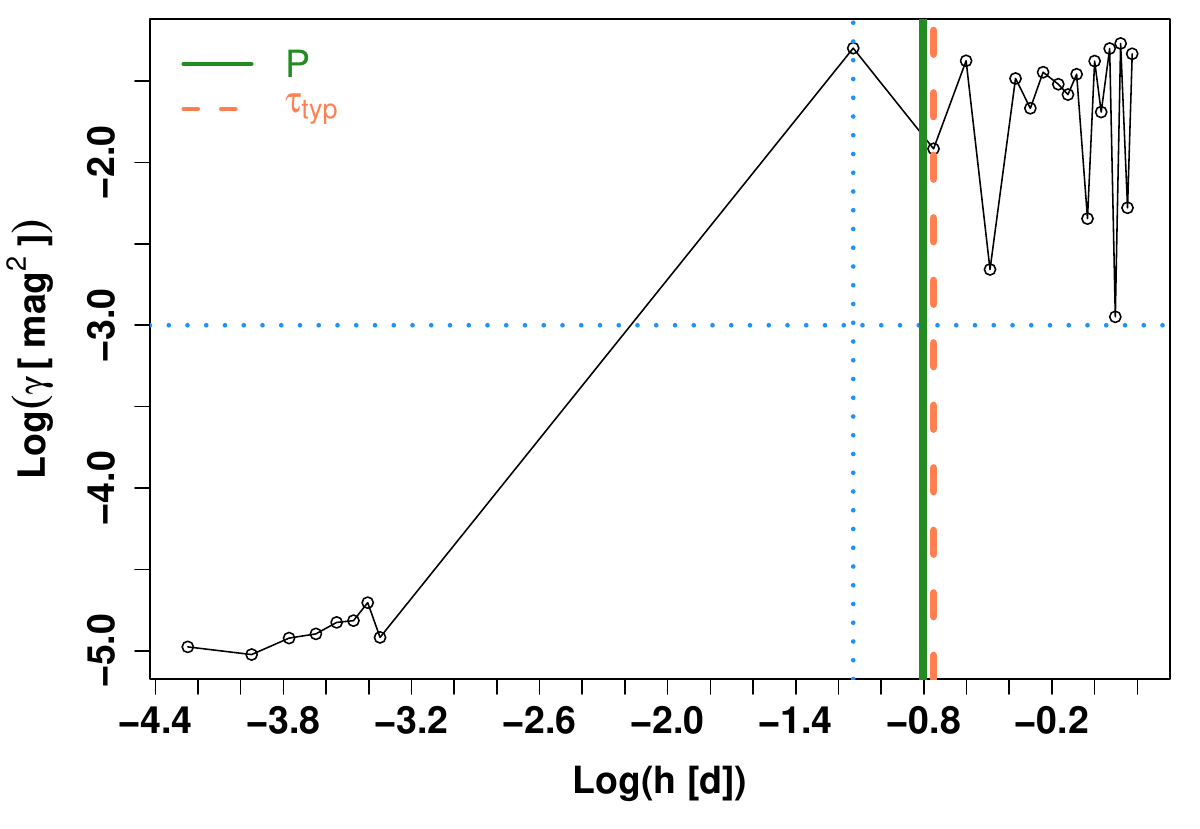}
\caption{Example of $\delta$ Scuti \textit{Gaia}-like light-curve (top), and corresponding unweighted observational variogram (bottom). The blue dotted lines indicate the detection threshold ($\gamma_{det} = 10^{-3} \,\mathrm{mag}^{2}$) and the associated detection timescale $\tau_{det}$. The green continuous line shows the simulation period $P$, and the orange dashed line corresponds to the typical timescale $\tau_{typ}$.}
\label{fig:dsct84Gaialike}
\end{figure}

\begin{figure}
\centering
\includegraphics[width=\columnwidth, page=1, trim={0 0 0 1.2cm}, clip=true]{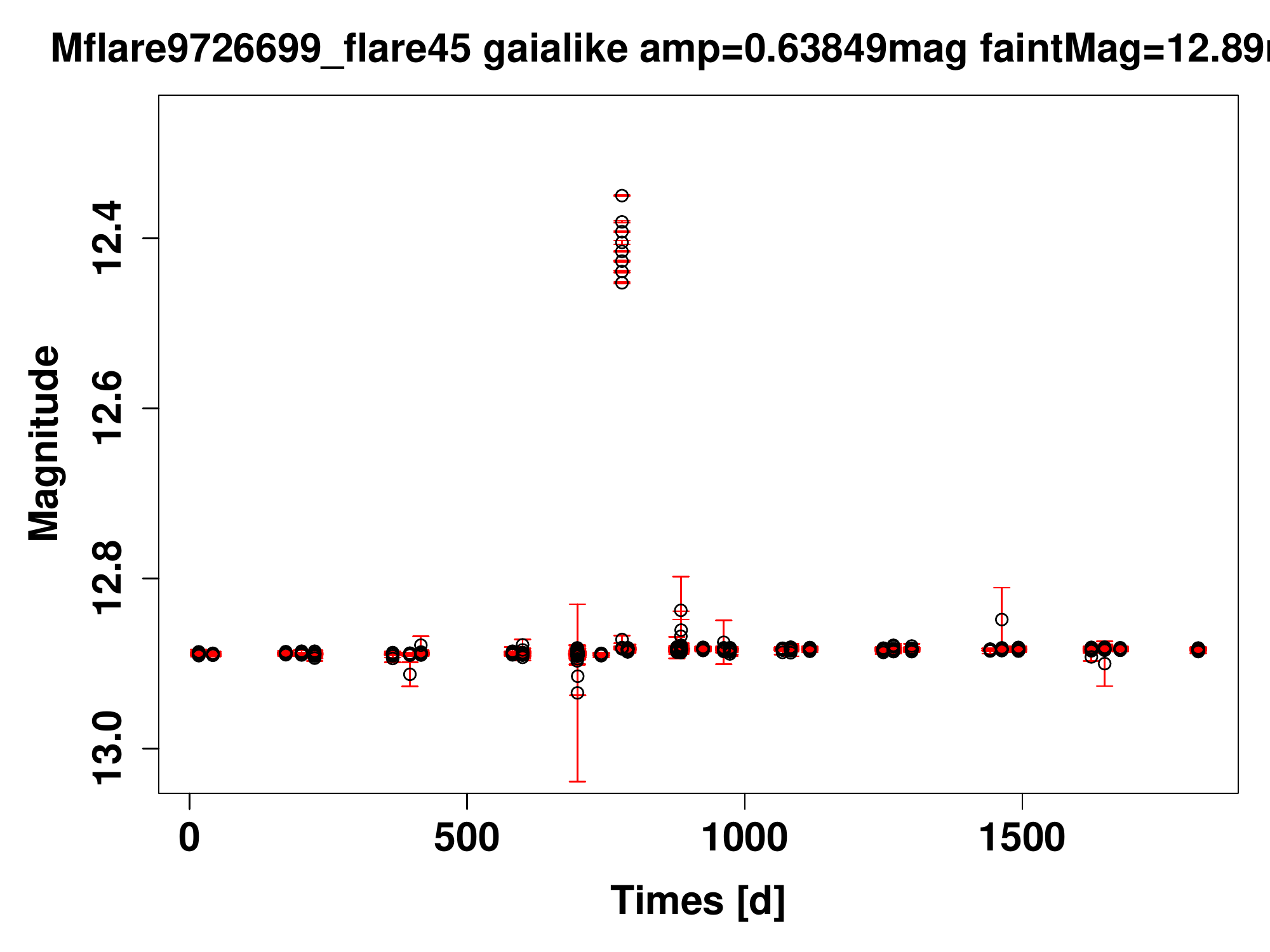}
\includegraphics[width=\columnwidth, page=2, trim={0 0 0 1.2cm}, clip=true]{Img_mnras/Mflare_9726699_flare45_gaialike_variogram_weighted_squareOfSumSigmaSquare_R_analysis.pdf}
\caption{Example of M flare \textit{Gaia}-like light-curve (top), and corresponding weighted observational variogram (bottom).}
\label{fig:MflareGaialike}
\end{figure}

\begin{figure}
\centering
\includegraphics[width=\columnwidth, page=1, trim={0 0 0 1.2cm}, clip=true]{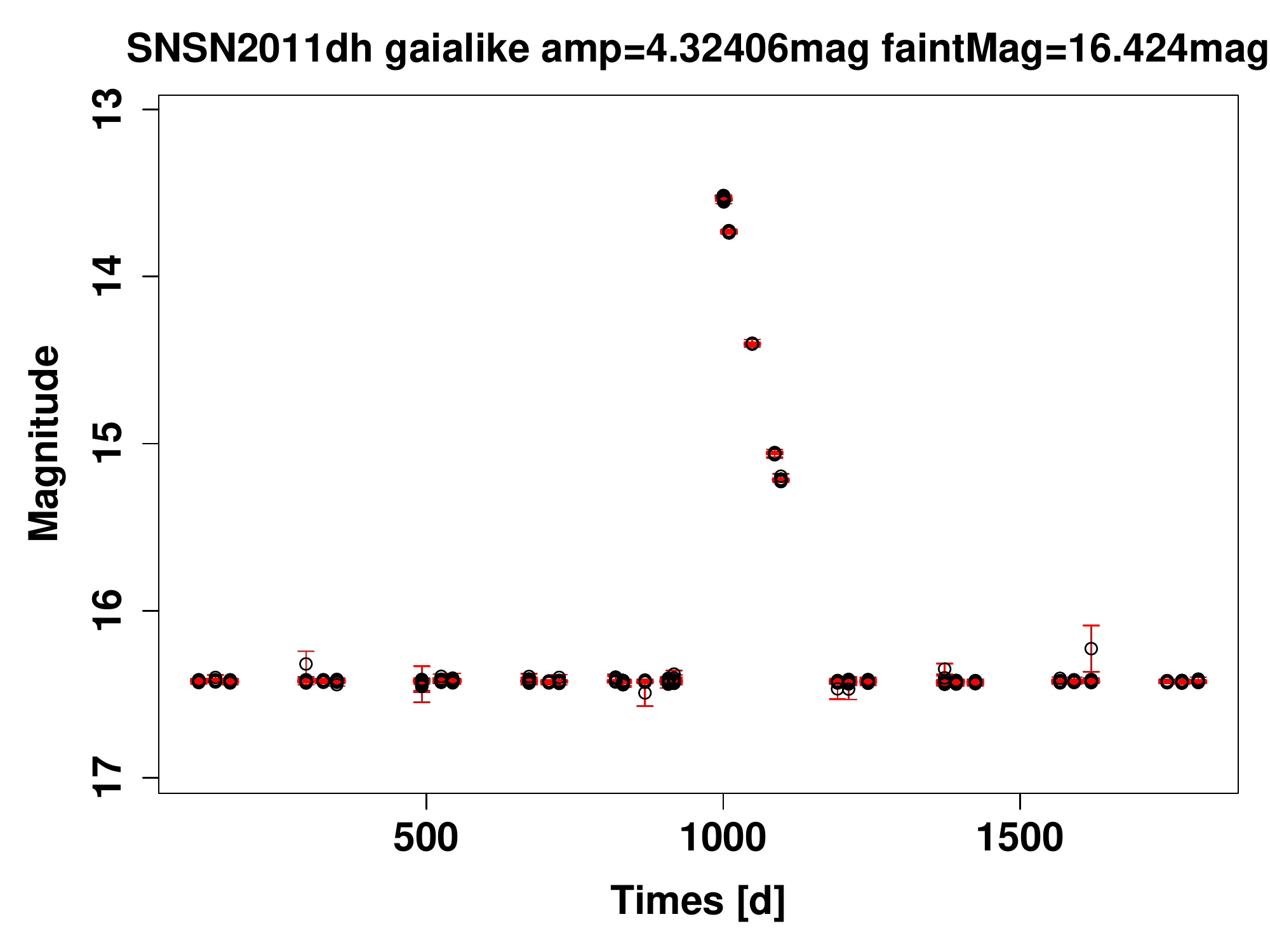}
\includegraphics[width=\columnwidth, page=2, trim={0 0 0 1.2cm}, clip=true]{Img_mnras/SN_SN2011dh_gaialike_variogram_weighted_squareOfSumSigmaSquare_R_analysis.pdf}
\caption{Example of SN \textit{Gaia}-like light-curve (top), and corresponding weighted observational variogram (bottom).}
\label{fig:SNGaialike}
\end{figure}


\subsection{Periodic variables}
\label{analysisPeriodicGaia}

First, we apply the same detection criterion as in Sect.~\ref{analysisContinuous}, with $\gamma_{det} = 10^{-3}\,\mathrm{mag}^{2}$ and the short timescale limit $\tau_{det} \leq 0.5\,$d. Figure \ref{fig:dsct84Gaialike} represents the \textit{Gaia}-like light-curve for the same simulated $\delta$ Scuti as for Fig. \ref{fig:dsct84continuous}. As can be seen, this source is also identified as short timescale variable in the \textit{Gaia}-like data set, this time with a detection timescale $\tau_{det} \approx 1\,$h$\,46\,$min instead of $9.1\,$min. In this case, the continuous detection timescale falls in the lag gap mentioned previously, thus in the \textit{Gaia} context the detection is pushed towards longer lags.

Among the 800 periodic variable simulations in the \textit{Gaia}-like data set, 647 (80.9\%) are flagged as short timescale variables, and 153 (19.1\%) are missed. Table \ref{tab:analysisContinuousVSGaialike} compares the detection results of the continuous and the \textit{Gaia}-like data sets for short period variables. Most of what is expected to be identified as short timescale variable candidate from the continuous data set is properly identified in the \textit{Gaia}-like data set. Similarly, most of what should not be flagged as short timescale variable from the ideal case is not flagged in the \textit{Gaia}-like context. However, a few percents of the simulated periodic sources that should not be flagged as short timescale variables from the ideal case are flagged in the \textit{Gaia}-like context. This is a consequence of the introduction of noise in the time series, thus increasing the measured variance level in the light-curve and pushing the variogram values above the detection threshold.

\begin{table}
\centering
\caption{Comparision of the detection results from variogram analysis between the continuous and the \textit{Gaia}-like data sets, with a single detection threshold $\gamma_{det} = 10^{-3}\,\mathrm{mag}^{2}$. We remind that the flagged sources}
\label{tab:analysisContinuousVSGaialike}
\includegraphics[width=0.9\linewidth]{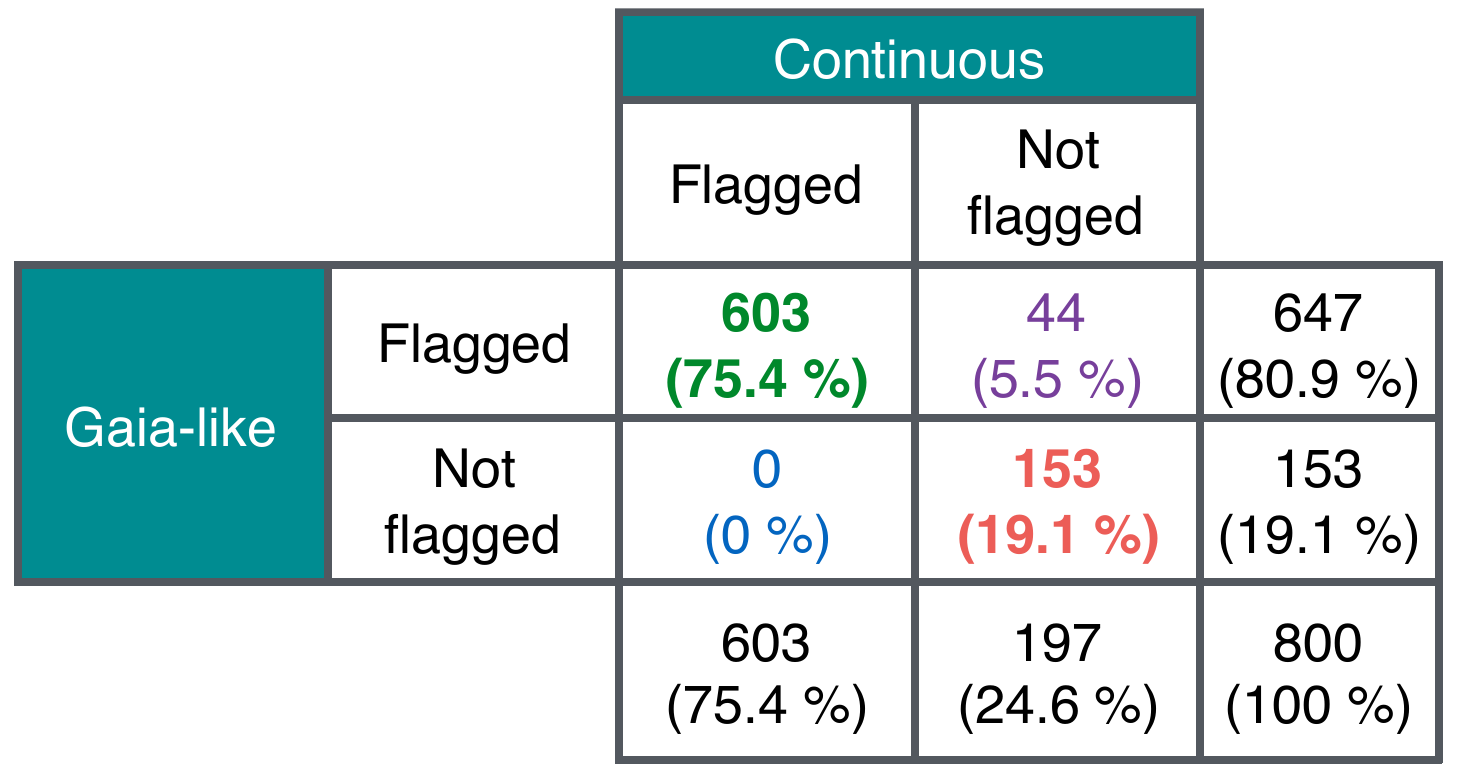}
\end{table}

\begin{figure}
\centering
\includegraphics[width=\columnwidth, trim={0cm 0.5cm 0.5cm 2cm}, clip=true]{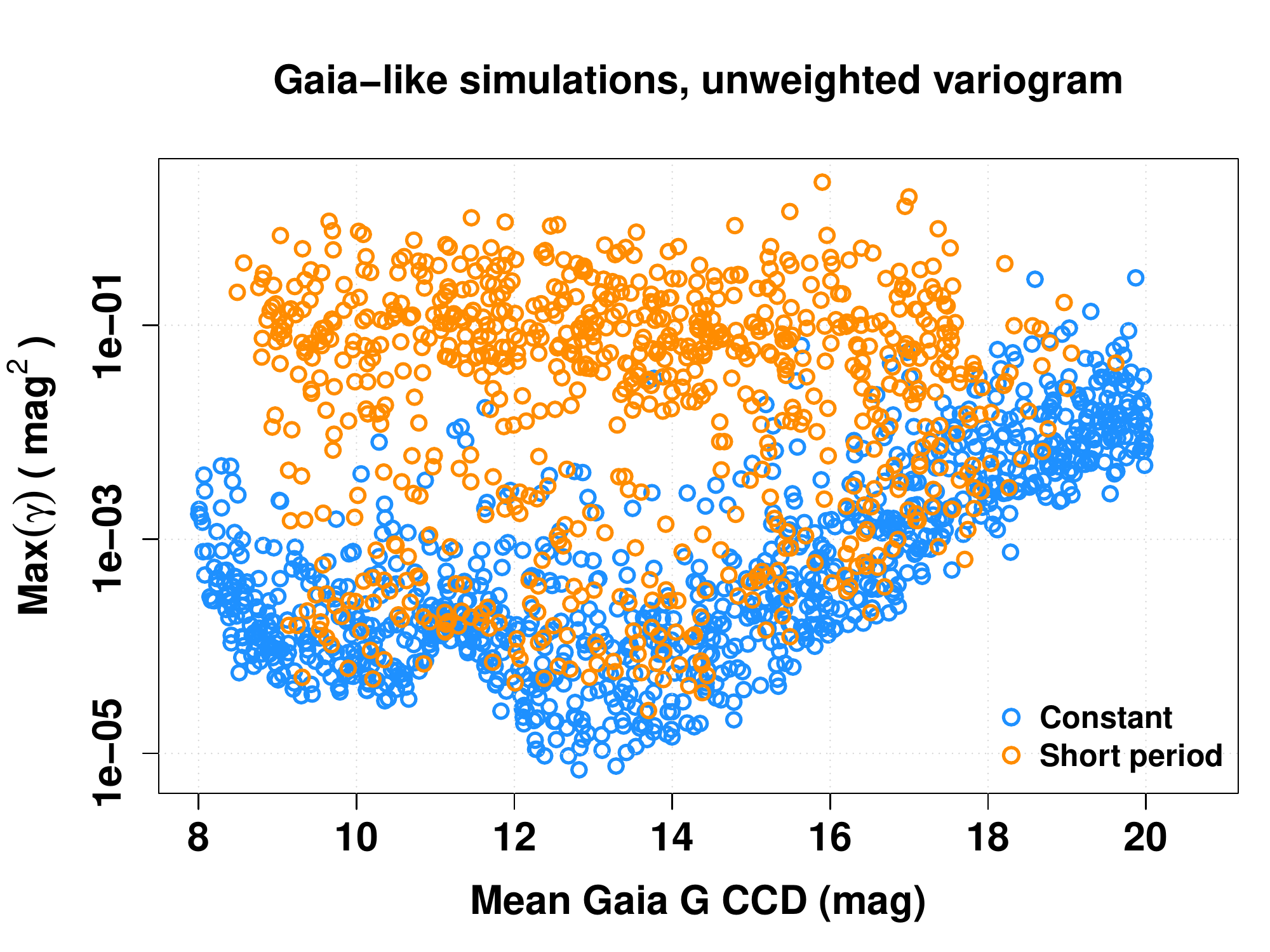}
\caption{Maximum variogram value (unweighted) as function of the mean \textit {Gaia} \textit{G} magnitude for the source, for the short timescale periodic variables and the constant sources of the \textit{Gaia}-like data set.}
\label{fig:maxVariogramVSMeanMagUnweightedWeighted}
\end{figure}

On the other hand, our short timescale detection criterion results in 350 of the 1000 simulated constant sources (35\%) that are flagged as short timescale candidates, which represents a huge contamination from constant sources. Figure \ref{fig:maxVariogramVSMeanMagUnweightedWeighted} shows the maximum variogram value for each simulated constant and short period source as function of their mean \textit{G} magnitude. As we can see, most of these detections correspond to constant sources fainter than $\sim 16-17\,$mag. For fainter sources, the noise measurement level for textit{Gaia} \textit{G} photometry becomes close to the variance limit fixed by the chosen value of $\gamma_{det}$. Hence $\gamma_{det} = 10^{-3}\,\mathrm{mag}^{2}$ is not adapted for faint sources. To minize the false positive rate without missing too many short period variables, a detection threshold function of the mean per-CCD magnitude of the source is required.\\
Nevertheless, whatever the magnitude, we see that, due to the presence of noise in the light-curves, the limit between short period variables and constant sources is unclear in the maximum variogram value - mean magnitude space. To better distinguish these two types of sources, we decide to adapt the variogram formulation using the weighted mean rather than the simple average. Considering a time series of magnitudes $(m_{i})_{i=1..n}$ with uncertainties $(\sigma_{i})_{i=1..n}$ observed at times $(t_{i})_{i=1..n}$, we define the \textit{weighted} variogram value for a lag h as
\begin{equation}
\gamma(h) = \frac{\sum_{i > j} w_{ij}(m_{j} - m_{i})^{2}}{\sum_ {i > j}w_{ij}} \,\mathrm{with}\,w_{ij} = \frac{1}{(\sigma_{i}^{2} + \sigma_{j}^{2})^{2}}
\end{equation}

\begin{figure*}
\centering
\includegraphics[width=0.7\linewidth, page=2, trim={0 0 0 2cm}, clip=true]{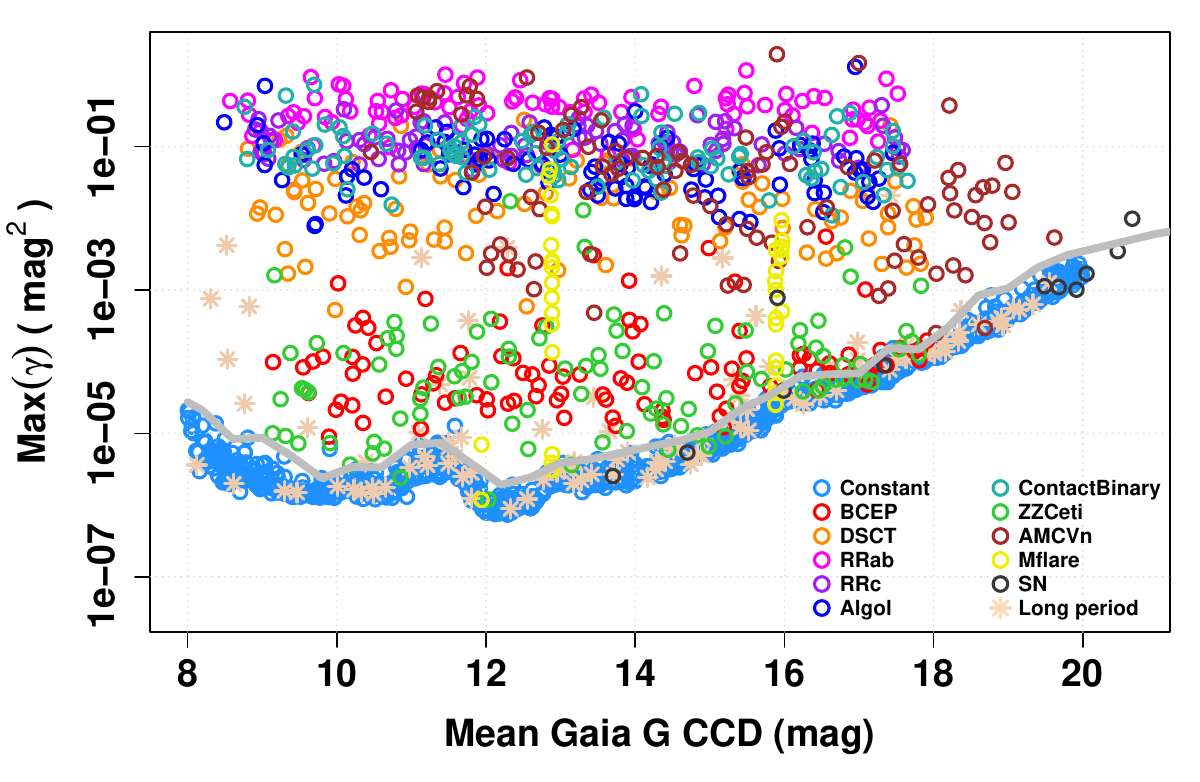}
\caption{Maximum weighted variogram value as function of the mean magnitude, for the short timescale periodic variables, the transients, the longer period variables and the constant sources of the \textit{Gaia}-like data set. The grey line shows the refined detection threshold depending on the mean magnitude of the source ($\gamma_{det} = \Gamma(\bar{G}_{CCD})$).}
\label{fig:maxVariogramVSMeanMagGaia}
\end{figure*}

Figure \ref{fig:maxVariogramVSMeanMagGaia} shows the maximum weighted variogram value as function of the mean magnitude of the source, for all the simulated short period variables and constant light-curves. Results for the simulated transient and longer period variables are also plotted for completness, but will be discussed later in Sect. \ref{analysisGaiaLike}. As we can see, this weighting scheme is very effective against false positives. Indeed, for constant stars, measurements with small uncertainties have very similar magnitudes (by definition of what a constant star is) and giving more weight to more precise magnitude paires means more weight to smaller magnitudes differences (thus smaller $\gamma$ values of constants), without affecting much the larger $\gamma$ of variable sources.
Note also that most of the improvement induced by weighting the variogram is done by the term $\sigma_{i}^2 + \sigma_{j}^2$, while the final square of this sum of uncertainties has been introduced to improve at the percent level.

With this new weighted variogram formulation, we can easily define a refined magnitude-dependent detection threshold: $\gamma_{det} = \Gamma(\bar{G}_{CCD})$ where $\Gamma$ is a piecewise linear function of the mean per-CCD magnitude of the source. The chosen $\Gamma(\bar{G}_{CCD})$ is represented by the grey continuous line in Figure \ref{fig:maxVariogramVSMeanMagGaia}.\\
When applying this new detection criterion, 768 of the 800 simulated periodic variables (96\%) are flagged as short timescale variables, and the false positive rate goes down to 0.1\%, which significantly improves the reliability of the variogram method. However, because hundreds of millions of sources observed by \textit{Gaia} will be investigated for short timescale variability, even a false positive rate of 0.1\% can result in a huge number of spurious detections. To reduce even more the variogram false positive rate, one possible solution is to use a more restictive detection criterion, with higher values of $\gamma_{det}$, e.g. $\gamma_{det} = 3\,\Gamma(\bar{G}_{CCD})$. With this definition, 717 of the 800 simulated periodic short timescale variables (89.6\%) are flagged short timescale, and none of the 1000 simulated constant sources is spuriously detected.
In both cases, most of the missed periodic variables are $\beta$ Cephei and ZZ Ceti stars, whose amplitudes are too low, compared to noise level at their magnitudes, to be properly identified. It also includes some AM CVn stars whose eclipses are not sampled enough to trigger detection from their variograms.

As mentioned previously, constant stars are not the only possible source of contamination of our short timescale variable candidate sample. Applying the variogram analysis to our set of longer period variables with the detection criterion $\gamma_{det} = \Gamma(\bar{G}_{CCD})$, 25 of the 100 simulated long period sources are flagged as short timescale variable candidates. With the more restrictive criterion $\gamma_{det} = 3\,\Gamma(\bar{G}_{CCD})$, the longer period contamination rate reduces to 14\%. These sources have periods shorter than $20\,$d or amplitudes greater than $0.25\,$mag. Though they are not short timescale variables per se, their global variation rate is high enough for their variogram to show significant values at lags below the $0.5\,$d short timescale limit we use. The occurence of such detections was somehow expected (see Sect. \ref{variogramMethod}), similarly to what we suspected regarding supernovae.
In the context of the \textit{Gaia} variability analysis, we could take advantage of the whole variability processing \citep{Eyer2015CU7,Eyer2017}, and remove this longer period contamination e.g. with the results of the classification module. We also plan to combine the variogram analysis with more accurate period search methods, such as Fourier based periodograms, which will also help distinguishing true short period variables from spurious longer ones. From a more general point a view, one way to limit the contamination of short timescale variable candidates by longer timescale variables could be to adopt a more restrictive definition of what short timescale variability is, e.g. with a lower detection timescale limit $\tau_{det} \leq 0.1\,$d, hence focusing on the fastest phenomena detected.

Table \ref{tab:variogramAnalysisGaiaLikeSummary} summarizes the results of the variogram analysis, for the weighted formulation, in terms of efficiency of short timescale variability detection, as well as false positive and longer period source contamination, with different detection criteria and short timescale limits. Detection results for the transient \textit{Gaia}-like simulations are detailed in Sect. \ref{analysisTransientGaia}.

\begin{table*}
\centering
\caption{Short timescale variability detection: results of the variogram analysis (with the weighted variogram formulation) for different detection criteria and different detection timescale limits. We remember that these percentages correspond to the ratio (number of sources of the considered class flagged as short timescale)/(total number of sources simulated for the considered class).}
\label{tab:variogramAnalysisGaiaLikeSummary}
\begin{tabular*}{\linewidth}{l p{0.25\linewidth} l p{0.25\linewidth} l p{0.25\linewidth} | p{0.25\linewidth} |}
\noalign{\smallskip}\hline\noalign{\smallskip}
\bf{Criterion} & \bf{Periodic short timescale recovery} & \bf{Constant contamination} & \bf{Longer period contamination}\\
\noalign{\smallskip}\hline\noalign{\smallskip}
$\gamma_{det} = 10^{-3}\mathrm{mag}^{2}$, $\tau_{det} \leq 0.5\,$d & 75.5\% & 4.6\% & 1\% \\
$\gamma_{det} = \Gamma(\bar{G}_{CCD})$, $\tau_{det} \leq 0.5\,$d & 96\% & 0.1\% & 25\% \\
$\gamma_{det} = \Gamma(\bar{G}_{CCD})$, $\tau_{det} \leq 0.1\,$d & 92.9\% & 0.1\% & 4\% \\
$\gamma_{det} = 3\,\Gamma(\bar{G}_{CCD})$, $\tau_{det} \leq 0.5\,$d & 89.6\% & 0\% & 14\% \\
$\gamma_{det} = 3\,\Gamma(\bar{G}_{CCD})$, $\tau_{det} \leq 0.1\,$d & 87.4\% & 0\% & 1\% \\
\noalign{\smallskip}\hline
\end{tabular*}
\end{table*}

For both the short period, longer period and constant classes, we have estimated the fraction of observed objects that would be identified as short timescale candidates by applying the variogram analysis to \textit{Gaia} photometry. But the real proportion between short period variables, longer period variables and constant sources over the sky is not the same as we simulated in our \textit{Gaia}-like data set. So how would our results be translated when turning to a more realistically distributed sample?
Among the over 1 billion celestial objects \textit{Gaia} will observe, we expect about 100 million sources showing variability, be it periodic, stochastic or temporally localized \citep{EyerCuypers2000}. In the Hipparcos Variability Annex \citep[see e.g.][]{Eyer1998}, about 30\% of the identified variables are periodic, and about 16\% of these periodic variables have periods shorter than $0.5\,$d. If we apply these percentages to the expected content of the \textit{Gaia} final catalogue, simply to get rough estimates, we end up with $\sim$ 4 million short period and $\sim$ 26 million longer period variables. Hence, from our variogram analysis results with $\gamma_{det} = \Gamma(\bar{G}_{CCD})$ and $\tau_{det} \leq 0.5\,$d, we would expect a total of 11.2 million sources flagged as short timescale candidates, including 3.8 million true short period variables, 6.5 longer period variables (possibly eliminated by post-processing), and 900,000 false positives. With $\tau_{det} \leq 0.1\,$d, we would get 5.6 million short timescale candidates, with 3.7 million true short period objects, 1 million longer period sources and 900,000 false positives, so a contamination of $\sim$ 18\% and 16\% from constant and longer period sources, respectively.

\begin{figure*}
\centering
\includegraphics[width=0.6\linewidth, page=4]{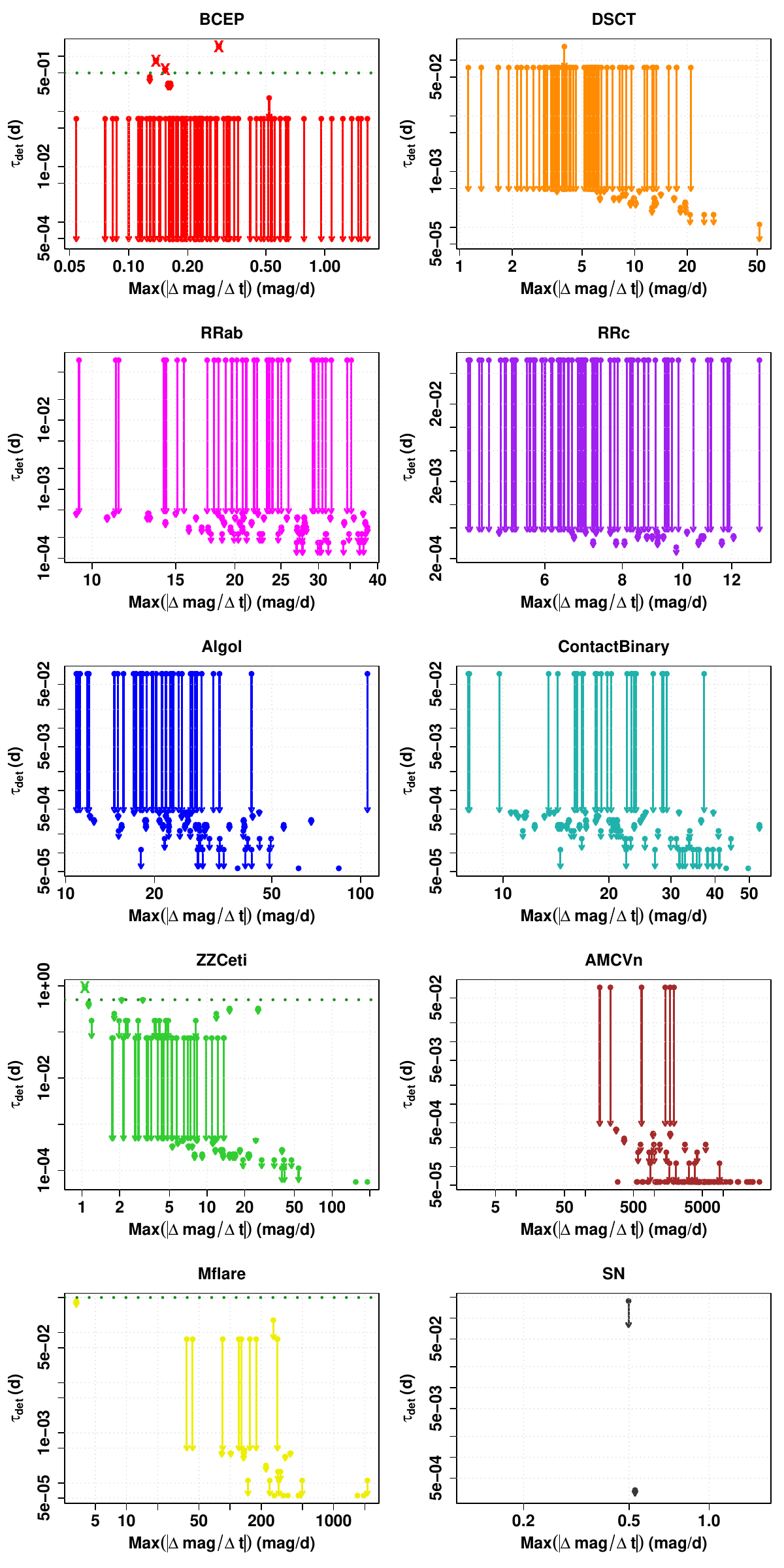}
\caption{Same as Fig. \ref{fig:tauDetMosaicContinuous}, for the \textit{Gaia}-like set, with the refined short timescale detection criterion $\gamma_{det} = \Gamma(\bar{G}_{CCD})$ and the detection timescale limit $\tau_{det} \leq 0.5\,$d. The dots indicate the detection timescale as it has been defined in Sect. \ref{variogramMethod}. The arrowheads indicate the lag just before $\tau_{det}$, i.e. the highest lag explored verifying $h < \tau_{det}$.}
\label{fig:tauDetMosaicGaia}
\end{figure*}

Figure \ref{fig:tauDetMosaicGaia} shows the $\tau_{det}$ distributions obtained with the detection criterion $\gamma_{det} = \Gamma(\bar{G}_{CCD})$, $\tau_{det} \leq 0.5\,$d, for each simulated periodic short timescale variable type. 
Note that, in the \textit{Gaia}-like context, the value of $\tau_{det}$ (dots in Fig. \ref{fig:tauDetMosaicGaia}) is rather an upper value of what the real detection timescale would be if we could explore more lags. Then, the lag just before $\tau_{det}$ (arrowheads in Fig. \ref{fig:tauDetMosaicGaia}), i.e. the highest lag explored veryfing $h < \tau_{det}$ is a lower limit for this real detection timescale. The two values ($\tau_{det}$ and the lag just before) define an interval where the ``true'' detection timescale would be if we could explore more lags between them.
We see that, in most cases, short timescale variability is detected at $\tau_{det} = 1\,$h$\,46\,$min, which is the duration between two successive \textit{Gaia}  FoV transits. However, $\tau_{det}$ can be as short as $10\,-\,20\,$s for the fastest $\delta$ Scuti, RR Lyrae, ZZ Ceti stars and eclipsing binaries. For AM CVn stars, which exhibit the highest variation rates among the periodic variable types simulated, $\tau_{det}$ can be as short as a few seconds. Note also that some of the simulated short period variables (e.g. some $\beta$ Cephei or ZZ Ceti stars) are detected, but with a detection timescale longer than the detection timescale limit, here $0.5\,$d.

\begin{figure}
\centering
\includegraphics[width=\columnwidth, page=1, trim={0 0 0 2cm}, clip=true]{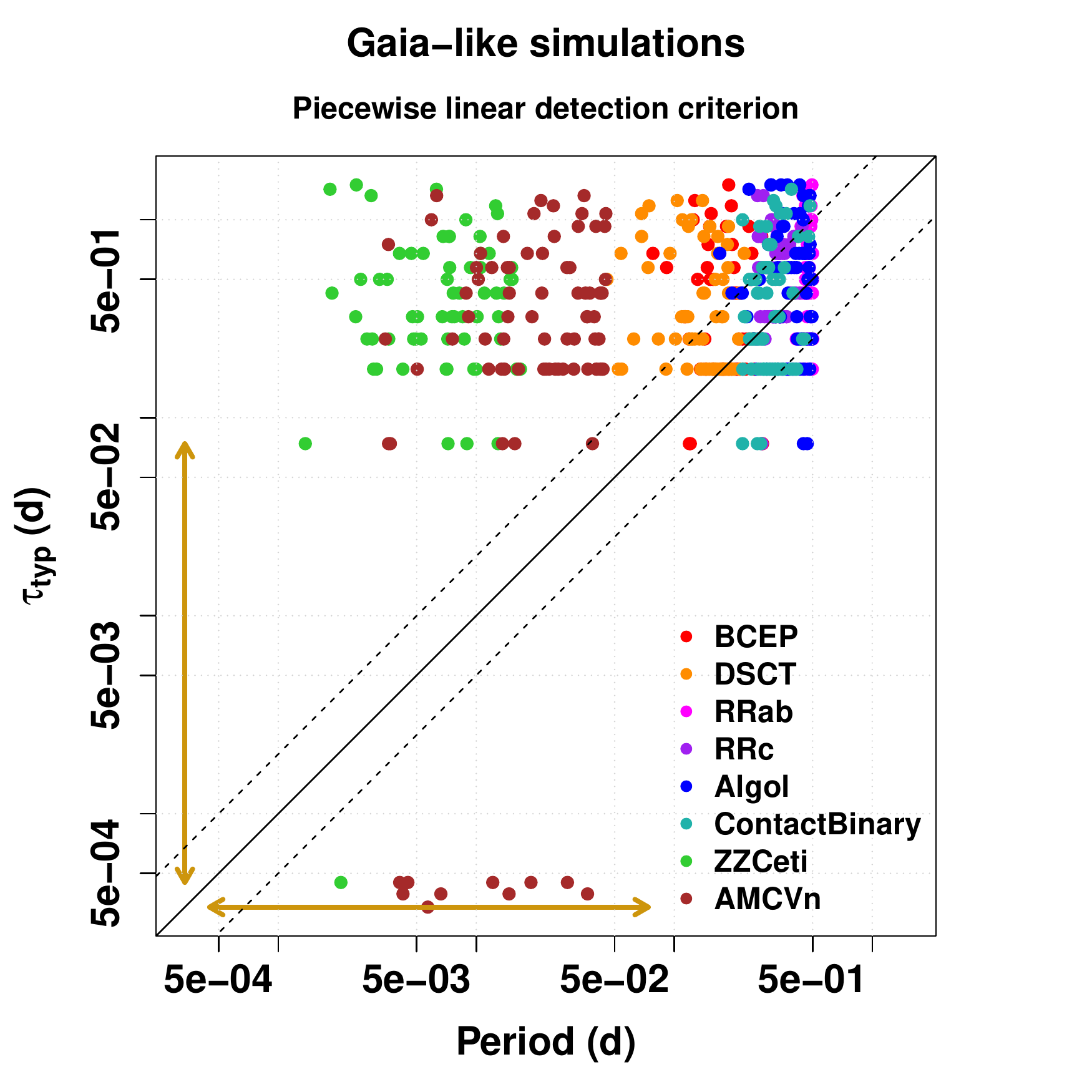}
\caption{Same as Fig. \ref{fig:tauTypVSPeriodContinuous}, for the \textit{Gaia}-like data set, with the refined short timescale detection criterion $\gamma_{det} = \Gamma(\bar{G}_{CCD})$ and $\tau_{det} \leq 0.5\,$d. The brown arrows indicate the \textit{Gaia} lag gap. The black dashed lines correspond to $\tau_{typ} = 2P$ and $\tau_{typ} = P/2$.}
\label{fig:tauTypVSPeriodGaia}
\end{figure}

Figure \ref{fig:tauTypVSPeriodGaia} represents the typical timescale estimated from the observational variograms of the 768 flagged short timescale periodic sources with the criterion $\gamma_{det} = \Gamma(\bar{G}_{CCD})$, $\tau_{det} \leq 0.5\,$d. For about 43\% of them, we have $P/2 \leq \tau_{typ} \leq 2P$. For 15\% of them, our method fails to provide a valid value of $\tau_{typ}$, because the maximum lag for which pairs of measurements can be formed is shorter that the variation period: this maximum lag is not high enough to get dips in the observational variogram, enabling to estimate $\tau_{typ}$. As one can see in Fig. \ref{fig:tauTypVSPeriodGaia}, a significant fraction of simulated sources have periods falling in the \textit{Gaia} lag gap, between $40\,$s and $1\,$h$\,46\,$min. For variable sources with a period within this interval, we do not expect to get a proper period estimate from the variogram. If we focus on the flagged simulations with a period outside the \textit{Gaia} lag gap (579 out of 768), the fraction of sources for which the typical timescale estimate fails remains at 15\%. For the other 85\% , $0.15P \leq \tau_{typ} \leq 15P$, and $\tau_{typ}$ recovers $P$ by a factor 2 in 56\% of the cases. Though it appears to be a quite low accuracy method for estimating periods, the variogram analysis on \textit{Gaia}-like photometry provides an order of magnitude for $P$.

In conclusion, with our refined short timescale detection criterion, the variogram method applied to the \textit{Gaia} per-CCD photometry should allow us to achieve a good recovery of short period variable candidates, with amplitudes down to a few mmag for the bright sources, and provide some information of their typical timescales, though period estimates are not expected to be very accurate. The fraction of constant sources observed and incorrectly flagged with this approach is reduced to a tenth of percent, and though the contamination from longer period variables is important, it should be significantly reduced by post-processing the initial list of short timescale candidates.

\subsection{Transient variables}
\label{analysisTransientGaia}

Among the 50 simulated transient events in the \textit{Gaia}-like data set, 33 M dwarf flares out of 38 are flagged as short timescale variables with the refined detection criterion and detection timescale limit at $0.5\,$d, as well as 2 supernovae out of 12. Figures \ref{fig:MflareGaialike} and \ref{fig:SNGaialike} show examples of flagged M flare and supernova, respectively.

Note that the 5 missed M flares are not detected at all ($\max(\gamma) < \gamma_{det}$). After visual inspection of the corresponding \textit{Gaia}-like light-curves, we find that either the flare is poorly sampled, or their amplitude is small relatively to the noise level in the time series. In both cases, variograms show no (or very little) evidence of variability. When turning to the two supernovae identified as short timescale candidates, we realize that one of them (SN2005hc) is very poorly sampled: because it is a faint source, with a quiescence magnitude of $25.4\,$mag and an amplitude around $7\,$mag, its measurements are expected to be most of the time below the \textit{Gaia} faint limit. Hence, its \textit{Gaia}-like light-curve contains only a few points sampling its brighter phase. Only few pairs of measurements can be formed, and the variogram values calculated are not really reliable. To limit such non-reliable detections, we impose a minimum number of 100 valid magnitude measurements for a \textit{Gaia} per-CCD time series to be investigated for short timescale variability. Note that this additional condition does not impact the results presented in Sect. \ref{analysisPeriodicGaia}.
As illustrated in Fig.\ref{fig:tauDetMosaicGaia}, detection timescales for M dwarf flares range between $10\,-\,20\,$s for the fastest simulated ones, and a few hours for the slowest ones. Regarding supernovae, if flagged as short timescale candidate then $\tau_{det}$ would be around a few hours (the point at $\tau_{det} \sim 10^{-4}\,$d corresponds to SN2005hc and should not be taken into account).

\begin{figure}
\centering
\includegraphics[width=\columnwidth, page=3, trim={0 0 0 2cm}, clip=true]{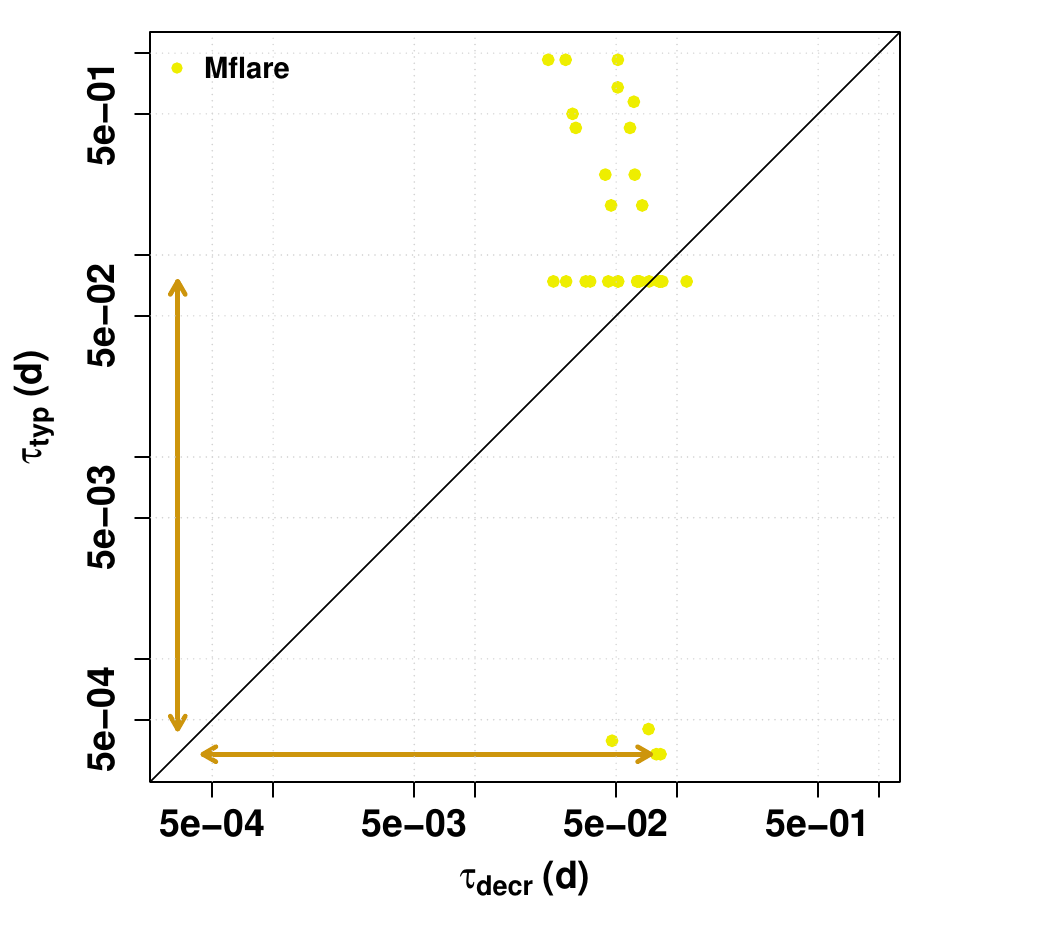}
\caption{Typical timescale $\tau_{typ}$ as function of the decrease duration, for the \textit{Gaia}-like transient simulations flagged as short timescale variables according to the refined short timescale detection criterion $\gamma_{det} = \Gamma(\bar{G}_{CCD})$ and $\tau_{det} \leq 0.5\,$d. The brown arrows indicates the \textit{Gaia} lag gap.}
\label{fig:tauTypVSDurationGaia}
\end{figure}

Figure \ref{fig:tauTypVSDurationGaia} represents the typical timescale estimates obtained for the 33 M dwarf flares identified as short timescale variables. At the moment, due to the  complexity of observational variograms of transients as compared to theoretical ones, we simply define the \textit{Gaia}-like typical timescale as the lag of the maximum variogram value, which should correspond to the decrease duration of the transient according to Sect. \ref{analysisTransientContinuous}. More refined techniques for pointing characteristic variogram features will be investigated in the future. Since the maximum lag explored is $1.5\,$d, which is much shorter than SNe decrease durations ($\sim$ a few tens of days), $\tau_{typ}$ is not expected to be relevant for this variable type. Concerning M flares, for 14 of the 33 flagged ones $\tau_{typ}$ recovers the decrease duration within a factor of 2, and for 23 out of 33 it recovers the decrease duration within a factor of 10. Hence, as for periodic variables, the variogram analysis gives an idea of the duration of transient events, though these estimates are not very accurate.

As mentioned in Sect. \ref{simulations}, until now we made sure that the simulated \textit{Gaia}-like transient light-curves sampled, at least partially, the event. However, in the real \textit{Gaia} context, the detectability of such phenomena will strongly depend on the times when measurements are taken relatively to the occurence of the event. Among all the fast flares, eclipses, explosions and occultations that will occur during the \textit{Gaia} lifetime, how many of them will be observed by the satellite and then detected as short timescale variable candidates? To assess this fraction, we generate 100 new \textit{Gaia}-like light-curves for each of the 50 transients considered, but this time without forcing to sample the event and randomizing its start time. Then we analyze each of them following our variogram approach for short timescale variable detection, i.e. at least $100$ valid per-CCD measurements, $\max(\gamma) \geq \gamma_{det}$ where $\gamma_{det} = \Gamma(\bar{G}_{CCD})$, and $\tau_{det} \leq 0.5\,$d.
Among the 3800 M flares simulated in this way, 678 are flagged as short timescale candidates, i.e. a fraction of about $18$\%. In the case of supernovae, 110 simulated light-curves out of 1200 trigger short timescale variability detection, which represent a fraction of $21$\%. Thus, with our variogram approach and the chosen detection criterion, we can expect to detect about $15\,-\,20$\% of the transient events occuring during the 5 years of the \textit{Gaia} mission.
Note that, if we fix the short timescale limit to $0.1\,$d instead of $0.5\,$d (for comparison purposes), we flag as short timescale candidates 8\% of the simulated M flares, and 8\% of the simulated supernovae.

\section{Conclusion}
\label{conclu}

In this work, by mean of extensive light-curve simulations, we showed that, with a specifically tailored detection criterion, the variogram method should enable a good recovery of short timescale variability, periodic or transient, from \textit{Gaia} per-CCD photometry, with a reduced fraction of observed constant sources resulting in false positives. Contamination for longer period variables is significant, and is essentially due to amplitude variations greater than $0.25\,$mag typically. It should be efficiently eliminated by post-processing involving both periodogram investigation and comparison with the other variability studies performed in the \textit{Gaia} DPAC context. We also showed that both contamination from constant and longer period variable sources can be limited by diminishing the short timescale limit fixed on $\tau_{det}$ from $0.5\,$d to $0.1\,$d, reducing the short timescale recovery rate only by a few percents.
Besides, this approach gives clues on the timescale(s) of the underlying variation. We saw that the typical timescale estimate provided by the variogram brings valuable information on the rapidity of the variation. In the case of periodic variability, it could be fruitfully combined with period search methods, e.g. to distinguish aliases from true periods.
Moreover, our analysis of simulated ideal short timescale variable light-curves allowed us to retrieve the specific timescale associated with a given variance level, for each of the short timescale variable type considered. This specific timescale gives indications on the observation cadence to adopt in the perspective of a ground-based photometric follow-up of such astronomical sources.
The next step of our study will be to re-invest the knowledge and understanding we acquired on the variogram analysis through simulations, to analyze real \textit{Gaia} per-CCD data and search for new short timescale variable candidates. Then, we will turn to the question of further classification and characterization of these \textit{Gaia} candidates, combining all the \textit{Gaia} data available (photometry in BP and RP, color, spectrum, parallaxes and proper motions) on the one hand, and complementing \textit{Gaia} output with observations from the ground on the other hand. We plan to explore the performance of machine learning methods, such as random forest, to assess the variable type of the selected candidates.

\section*{Acknowledgements}
This work has made use of data from the ESA space mission \textit{Gaia}, processed by the \textit{Gaia} Data Processing and Analysis Consortium (DPAC).

Some of the data presented in this paper were obtained from the Mikulski Archive for Space Telescopes (MAST). STScI is operated by the Association of Universities for Research in Astronomy, Inc., under NASA contract NAS5-26555. Support for MAST for non-HST data is provided by the NASA Office of Space Science via grant NNX09AF08G and by other grants and contracts. This paper includes data collected by the Kepler mission. Funding for the Kepler mission is provided by the NASA Science Mission directorate.

We acknowledge with thanks the variable star observations from the AAVSO International Database contributed by observers worldwide and used in this research.

This research has made use of the CfA Supernova Archive, which is funded in part by the National Science Foundation through grant AST 0907903.

We gratefully acknowledge Mihaly Varadi for kindly providing us ZZ Ceti star light-curve simulations, that we used to construct our short timescale variable templates.
\bibliographystyle{mnras}
\bibliography{variogramme_draft_v1_mybib}

%

\end{document}